\documentclass[a4paper,superscriptaddress,twocolumn,prb,showkeys,preprintnumbers,floatfix,showpacs]{revtex4-1}

\usepackage{epsfig}
\usepackage{amsmath}
\usepackage{amssymb}

\usepackage[colorlinks,bookmarks=false,citecolor=blue,linkcolor=red,urlcolor=blue]{hyperref}

\begin{document}

\title{Ground-State Phase Diagram of the 1D t-J model}

\author{Alexander Moreno}
\affiliation{Institut f\"{u}r Theoretische Physik III, Universit\"{a}t Stuttgart, Pfaffenwaldring 57, 70550 Stuttgart, Germany}
\author{Alejandro Muramatsu}
\affiliation{Institut f\"{u}r Theoretische Physik III, Universit\"{a}t Stuttgart, Pfaffenwaldring 57, 70550 Stuttgart, Germany}
\affiliation{Kavli Institute for Theoretical Physics, University of California, Santa Barbara, CA 93106}

\author{Salvatore R. Manmana}
\affiliation{JILA, University of Colorado and NIST, and Department of Physics, University of Colorado, Boulder, CO 80309-0440, USA}
\preprint{NSF-KITP-10-162}

\date{\today}

\begin{abstract}
We examine the ground-state phase diagram of the t-J model in one dimension by means of the Density Matrix Renormalization Group. 
This model is characterized by a rich phase diagram as a function of the exchange interaction $J$ and the density $n$, displaying Luttinger-liquid (LL) behavior both of repulsive and attractive (i.e.\ superconducting) nature, a spin-gap phase, and phase-separation.
The phase boundaries separating the repulsive from the attractive LL phase as $J$ is increased, and also the boundaries of the spin-gap region at low densities, and phase-separation at even larger $J$, are determined on the basis of correlation functions and energy-gaps. In particular, we shed light on a contradiction between variational and renormalization-group (RG) results about the extent of the spin-gap phase, that results larger than the variational but smaller than the RG one. Furthermore, we show that the spin gap can reach a sizable value ($\sim 0.1 t$) at low enough filling, such that preformed pairs should be observable at temperatures below these energy scales. No evidence for a phase with clustering of more than two particles is found on approaching phase separation.     
\end{abstract}

\maketitle
\section{Introduction}
The t-J model constitutes together with the Hubbard model a paradigm for the theoretical description of high temperature superconductors (HTS) since its derivation by Zhang and Rice \cite{zhang98} from a three-band Hubbard (spin-fermion) model describing the copper-oxide planes present in HTS. 
It can also be derived in second order perturbation theory around $U=\infty$, where $U$ is the strength of the interaction in the Hubbard model \cite{chao77}. Its Hamiltonian is 
\begin{eqnarray}
H & = & -t \sum _{\langle i, j  \rangle \atop \sigma}  
\left( f_{i,\sigma}^\dagger f^{}_{j,\sigma} + \mbox{h.c.} \right) 
\nonumber \\ & &
+ J \sum _{\langle i, j \rangle} \left( {\vec S}_i \cdot {\vec S}_{j} - \frac{1}{4} n_i n_{j}   \right),
\label{Hamiltonian}
\end{eqnarray}
where the operator $f_{i,\sigma} ^\dagger$ ($f_{i,\sigma}$) creates (destroys) a fermion with spin $\sigma= \uparrow, \downarrow$ on the site $i$. They are not canonical fermionic operators since they act on a restricted Hilbert space without double occupancy.  
${\vec S}_i = f^\dagger_{i,\alpha} \vec \sigma_{\alpha \beta} f^{}_{i,\beta}$ is the spin operator and $n_i= f_{i,\sigma}^\dagger f^{}_{i,\sigma} $ is the density operator. In all expressions a summation over repeated indices is understood. Furthermore, $\langle i, j \rangle$ denotes nearest neighbor bonds. 

While the main interest on this model resides on its two-dimensional (2D) realization, it presents already in one dimension (1D) a number of very interesting features. In contrast to the 1D repulsive Hubbard model, where only a Luttinger liquid (LL) phase for density $n \neq 1$ and an insulating phase for $n=1$ are present, the t-J model possesses a rich phase diagram, as shown first by    
M.\ Ogata {\em et al}. \cite{ogata91}. 
Interestingly, the phases display a correspondence to the ones present in  HTS, like superconductivity, spin gap and phase separation, albeit for values of $J/t$ outside the range pertaining to HTS. 
However, since unbiased results for the 2D model, that up to now could only be obtained by exact diagonalization \cite{dagotto94} or by density matrix renormalization group (DMRG) \cite{white98}, did not yet conclusively shed light on the different phases in the thermodynamic limit, the 1D version presents a possibility of gaining insight into exotic phases like the spin-gap region.

A further motivation for achieving an accurate determination of the phase diagram of this model is the possibility of realizing it with ultra-cold fermionic quantum gases\cite{eckardt10,reypriv}. For parameters aiming at an emulation of HTS, still cooling techniques have to be implemented to reach the relevant energy scales ($\sim J/10$, with $J \sim 0.3 t$). On the other hand, as discussed in the following, the spin-gap phase appears here at much larger values of $J$ ($J \sim 2.5 t$), and hence this non-trivial phase is certainly much more accessible for experiments with degenerate quantum gases.

The 1D t-J model has been solved exactly only for $J/t \rightarrow 0$, where it is equivalent to the $U \rightarrow \infty$ Hubbard model \cite{schulz90,kawakami90,frahm90}, and at the supersymmetric point $J=2t$ \cite{bares90, bares91}. In both cases the model behaves as a LL \cite{giamarchi04, haldane81, haldane81b, solyom79, emery79}. For very large $J/t$ the attractive interaction dominates against the kinetic energy and the system phase separates into hole rich regions and antiferromagnetic islands. Although the first phase diagram appeared almost twenty years ago, there are still issues to be clarified like the 
boundaries of the spin-gap phase. In previous studies the existence of a spin gap has been deduced from exact diagonalization (ED) of small systems \cite{ogata91}, variational \cite{chen93}, or projection \cite{helberg93} methods, transfer-matrix renormalization group (TMRG),\cite{SirkerKluemper2002}
and by a combination of renormalization group (RG) arguments with ED
\cite{nakamura97}. In particular the last study found a much larger region in density for the spin-gap phase.

A main goal of the present work is to achieve a precise determination of the phase diagram, by performing finite-size extrapolations to the thermodynamic limit for the correlation functions and energy differences relevant for the different phases. The results are obtained using the DMRG method \cite{white92,white93,schollwoeck05}, on lattices with up to $L = 200$ sites. We have to choose values of $n$ and $L$ such that
the total particle number
$N=n L$ is an integer number. In most of the results we extrapolate to the thermodynamic limit using system sizes $L=40$, 80, 120, 160, and 200. These values of $L$ allow us a discretization in density of $\Delta n=0.05$ which is consistent with $N$ being an even integer, and hence, the ground-state corresponds to $S_{tot}^{z} = 0$. All the results were obtained using at least $200$ DMRG vectors, 4 sweeps, and a discarded weight of $10^{-8}$. This translates into errors in energy of the order of 10$^{-8}$  and in correlation functions of the order of 10$^{-5}$ at the largest distances.

\section{Phase diagram\label{Sec:PhaseDiagram}}

The phase diagram for the 1D t-J model obtained by Ogata et al.\ \cite{ogata91} is based on ED on systems with up to 16 lattice sites. They found three phases: repulsive LL phase (metal), attractive LL phase (superconductor) and phase-separation. At that time they suspected the existence of a spin gap at low density but they could not prove its existence due to the limitation to small system sizes. Posterior works \cite{chen93, helberg93} found evidences of the spin gap using variational methods. However, Nakamura {\em et al}. \cite{nakamura97} found based on an RG analysis that the spin-gap region is larger than expected. Here we present results from a direct measurement of the spin gap and an extrapolation to the thermodynamic limit. Details for it will be discussed in Sec.\ \ref{spin-gap}.   

\begin{figure}[th!]\relax
\centerline{\includegraphics[width=3in]{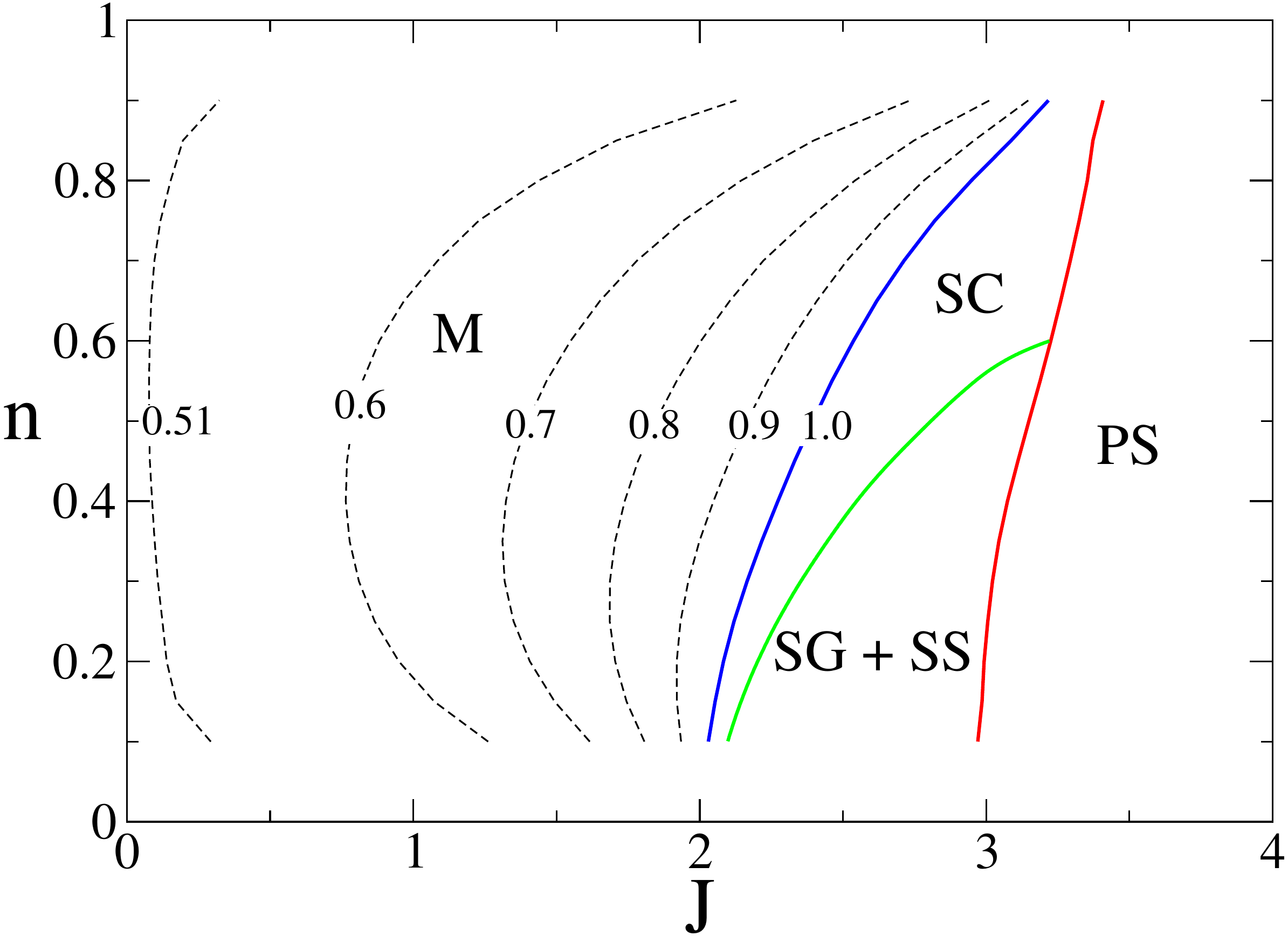}}
\caption{(color online). Phase diagram of the 1D t-J model from DMRG for densities $0.1 \leq n \leq 0.9$ and in the range $0 < J \leq 4$, where  we set $t=1$. 
$n=N/L$ is the electronic density ($N$ is the total number of particles and $L$ the number of lattice sites). Four phases are present: a metallic phase (M) or repulsive Luttinger liquid, a gapless superconducting (SC) phase, a singlet-superconducting phase with spin gap (SG + SS), and phase-separation (PS). 
The number given to each line stands for the value of the Luttinger parameter $K_\rho$.}
\label{fig1}
\end{figure}
Our results can be summarized in the phase diagram shown in Fig.\ \ref{fig1}. We obtained four phases:
a metallic phase (M) or repulsive LL, a gapless superconducting (SC) region, 
a singlet-superconducting phase with spin gap (SG + SS), and phase-separation (PS), where the system separates into a hole-rich and an electron-rich part. 
The number given to each line stands for the value of the Luttinger parameter $K_\rho$, with $K_\rho < 1.0$ in M and $K_\rho > 1.0$ in both superconducting phases. The determination of $K_\rho$ will be discussed in the subsequent sections. 

In order to characterize the phases and to find the boundaries between them, we calculated directly the energy gap to triplet excitations and measure the density-density correlation functions,
\begin{equation}
N_{ij} = \langle n_i n_j \rangle - \langle n_i \rangle \langle n_j \rangle,
\label{e1}
\end{equation} 
the spin-spin correlation function,
\begin{equation}
S_{ij} = \langle S_i^z S_j^z \rangle,
\label{e2}
\end{equation} 
the pairing correlation function 
\begin{equation}
P_{ij} = \langle \Delta_i^\dagger \Delta_j \rangle,
\label{e3}
\end{equation} 
where
\begin{equation}
\Delta_i^\dagger = \frac{1}{\sqrt{2}} (f_{i,\downarrow} ^\dagger f_{i+1,\uparrow} ^\dagger - f_{i,\uparrow} ^\dagger f_{i+1,\downarrow} ^\dagger)
\label{e3.1}
\end{equation}
for singlet pairing, and
\begin{equation}
\Delta_i^\dagger = f_{i,\uparrow} ^\dagger f_{i+1,\uparrow} ^\dagger 
\label{e3.2}
\end{equation} 
for triplet pairing. Finally, we also considered the one-particle Green's function 
\begin{equation}
G_{ij} ^\sigma = \langle f_i f_j^\dagger \rangle.
\label{e4}
\end{equation} 
The corresponding structure factors are obtained by Fourier transformation,
\begin{equation}
X(k) = \frac{1}{L} \sum _{i,j=1} ^L e^{ik(x_i-x_j)} X_{ij},
\label{e5}
\end{equation} 
Although the systems considered lack translational invariance due to open boundary conditions, for the large system sizes considered here, we could not observe any artifact introduced by this procedure.
  
\subsection{Metallic phase\label{MetallicPhase}}

In order to characterize this phase, we compute the Luttinger parameter $K_\rho$, with 
$K_\rho < 1$ ($K_\rho > 1$) for a  repulsive (attractive) interaction, and $K_\rho = 1$ for the free case.
In oder to obtain $K_\rho $, we consider the limit $k \rightarrow 0$, where the structure factor for the density correlations displays a linear behavior with a slope proportional to $K_\rho$\cite{clay99,ejima05, giamarchi04},
\begin{equation}
N(k) \rightarrow K_\rho |k|a/\pi \text{ for } k \rightarrow 0,
\label{e6}
\end{equation} 
that results from Fourier transforming the first term in Eq.\ (\ref{ncorr}) below.
Here $a$ is the lattice constant (we set $a=1$). 
   \begin{figure}[th!]\relax
\centerline{\includegraphics[width=3in]{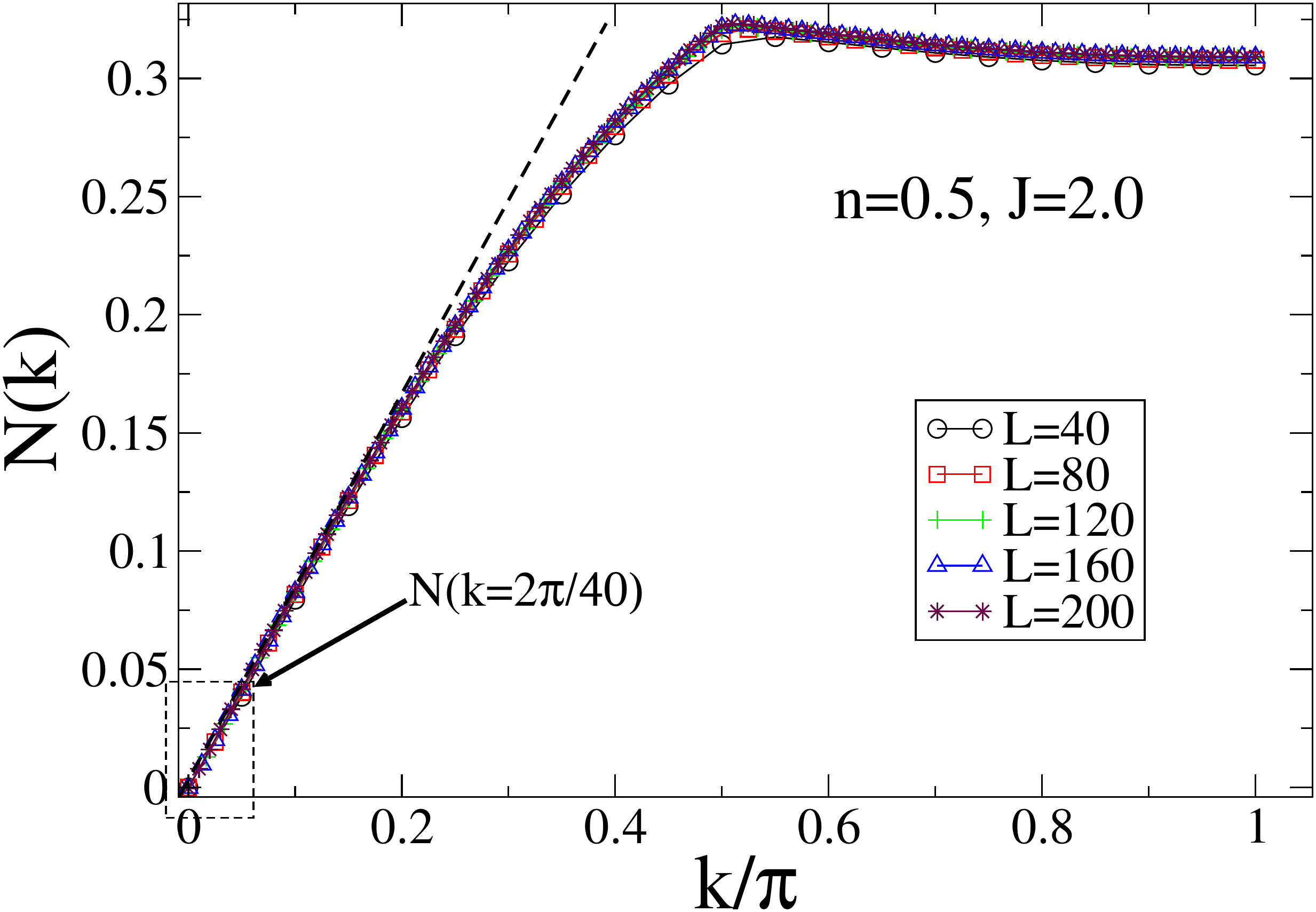}}
\caption{(color online). Structure factor $N(k)$ of the density-density correlation function for $n=0.5$, $J=2.0$ and $L=40$, 80, 120, 160, and 200.}
\label{fig2}
\end{figure}
Figure \ref{fig2} shows $N(k)$ for $n=0.5$, $J=2.0$ and $L=40$, 80, 120, 160, and 200. We observe a clear linear behavior for small $k$, with $N(k=0)=0$ due to the conservation of the total particle number.

\begin{figure}[th!]\relax
\centerline{\includegraphics[width=3in]{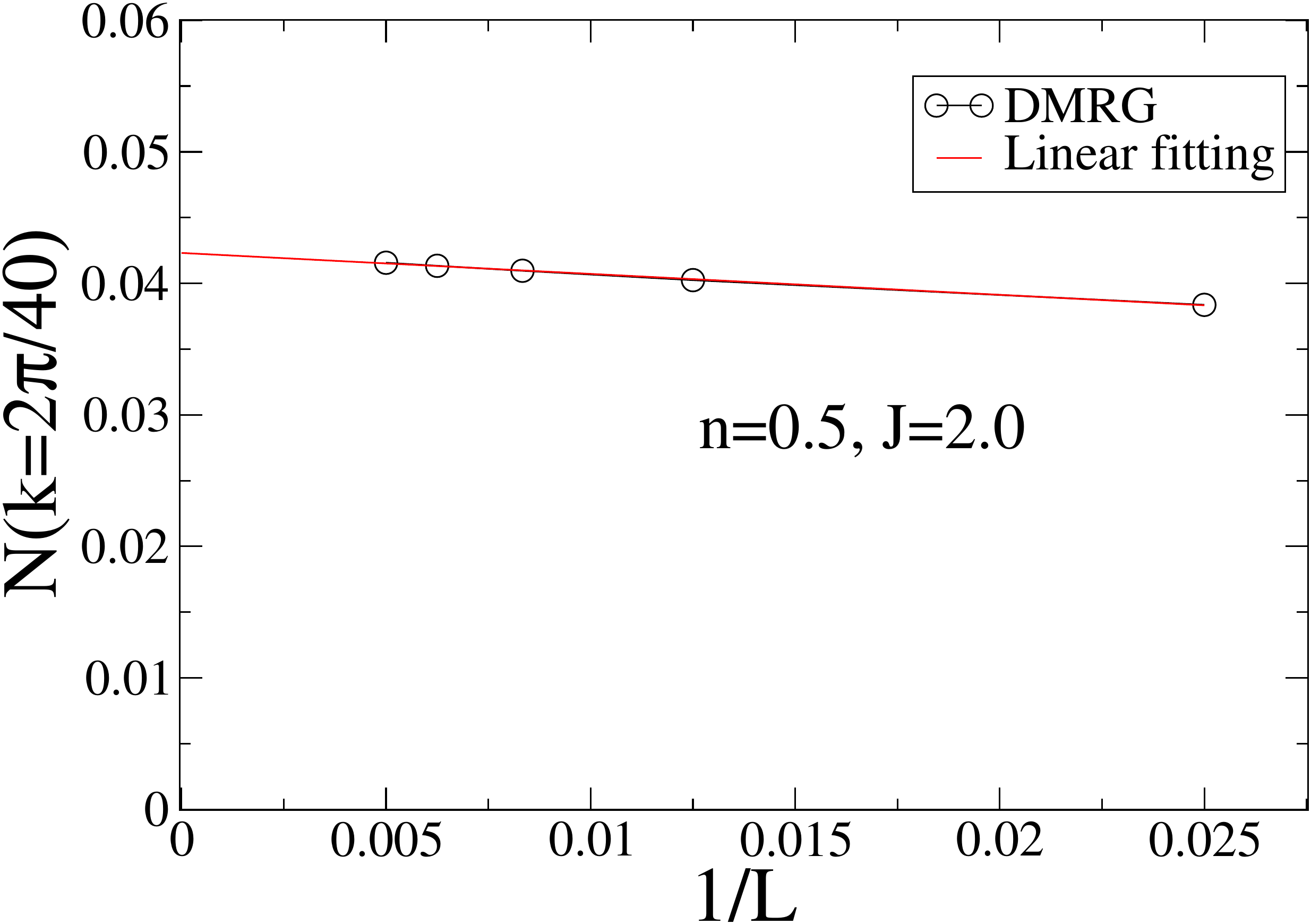}}
\caption{(color online). Extrapolation to the thermodynamic limit of $N(k=2\pi/40)$ for $n=0.5$ and $J=2.0$.}
\label{fig3}
\end{figure}

Although $N(k)$ appears to be almost independent of the lattice size, a more precise value of the slope is obtained by extrapolating the value of $N(k)$ at the point $k=2\pi/40$, that is the smallest wavevector in our smallest system, to $L\rightarrow \infty$. Using this last value and $N(k=0)=0$ we obtain the slope and then we can extract $K_\rho$ in the thermodynamic limit using Eq.\ (\ref{e6}). This extrapolation is shown in Fig.\ \ref{fig3}. 

\begin{figure}[ht!]\relax
\centerline{\includegraphics[width=3in]{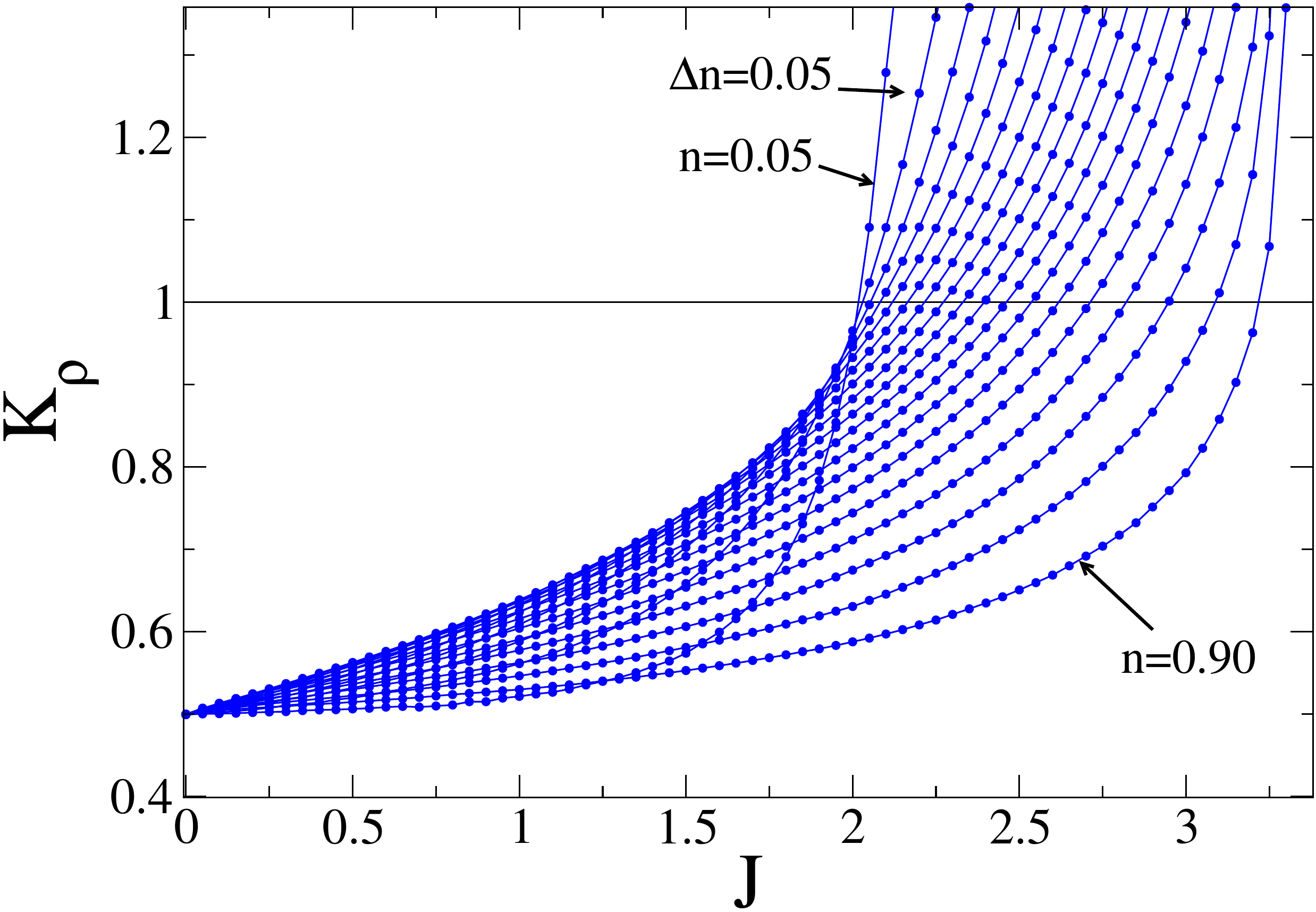}}
\caption{(color online). $K_\rho$ as function of J for different densities $n$.}
\label{fig4}
\end{figure}

We repeated this procedure for different values of $n$ and $J$. $K_\rho$ as function of $J$ for different densities $n$ is plotted in Fig.\ \ref{fig4}. Note that $K_\rho \rightarrow 0.5$ when $J \rightarrow 0$ for all densities, which is in agreement with the results obtained for the $U/t \rightarrow \infty$ Hubbard model \cite{schulz90}. It can also be observed that, for $K_\rho > 1$, $K_\rho$ increases quite fast with the interaction constant $J$ and it actually should diverge in the phase-separated region. 

\begin{figure}[th!]\relax
\centerline{\includegraphics[width=3in]{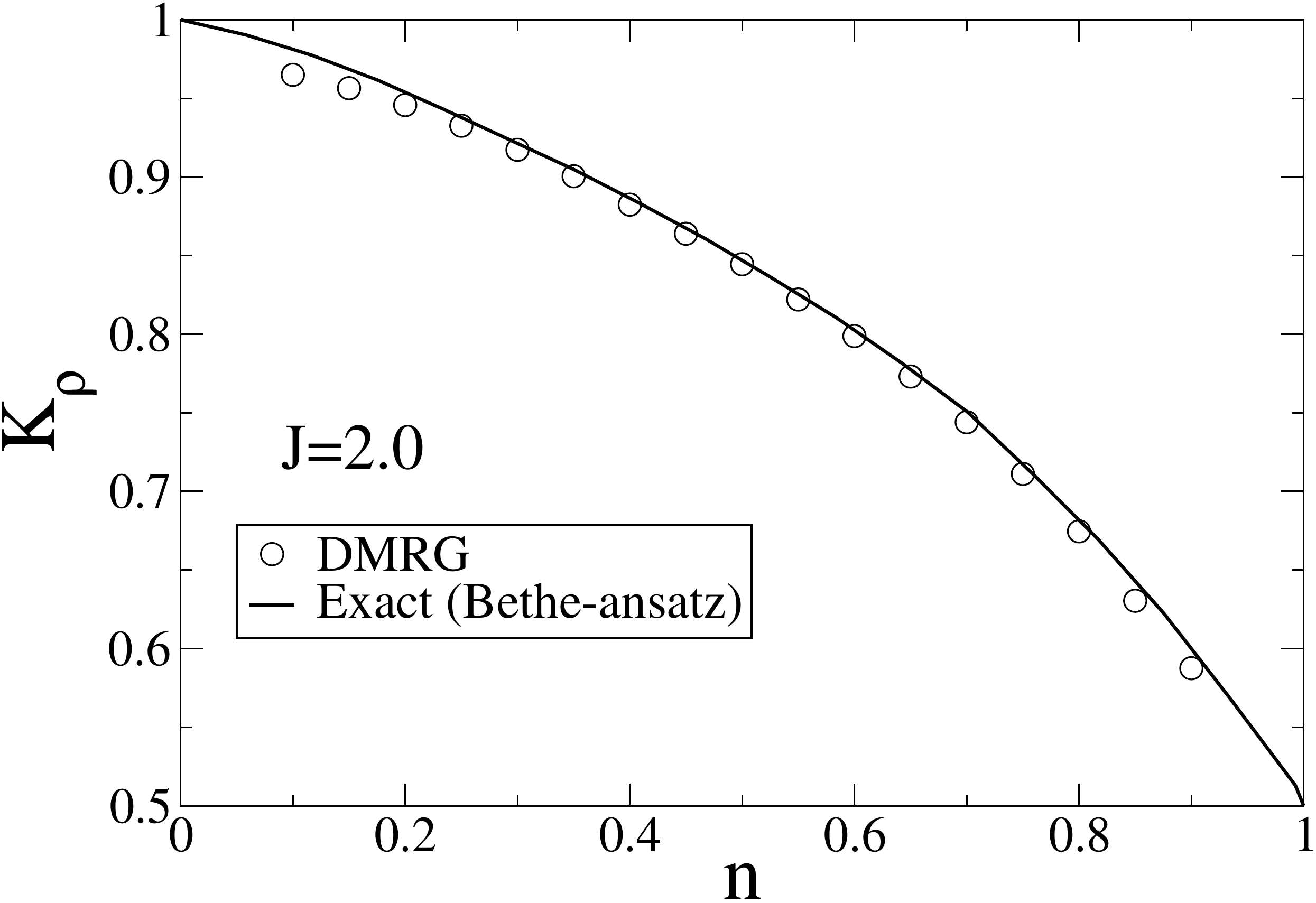}}
\caption{$K_\rho$ as function of the density $n$ for $J=2$ (supersymmetric point).}
\label{Kp_Bethe}
\end{figure}

The critical exponents $K_\rho$ at the supersymmetric point $J=2$ and for all densities were exactly obtained by means of the Bethe ansatz \cite{kawakami90b}. In Fig.\ \ref{Kp_Bethe} we compare our DMRG results with this exact solution,  and observe a very good agreement between both.
The deviations at very low (very high) densities point to the necessity of having larger systems for such very dilute cases, with a rather small number of particles (holes) to be able to properly describe a phase. They lead however, to barely noticeable shifts in the phase diagram. 

\begin{figure}[th!]\relax
\centerline{\includegraphics[width=3in]{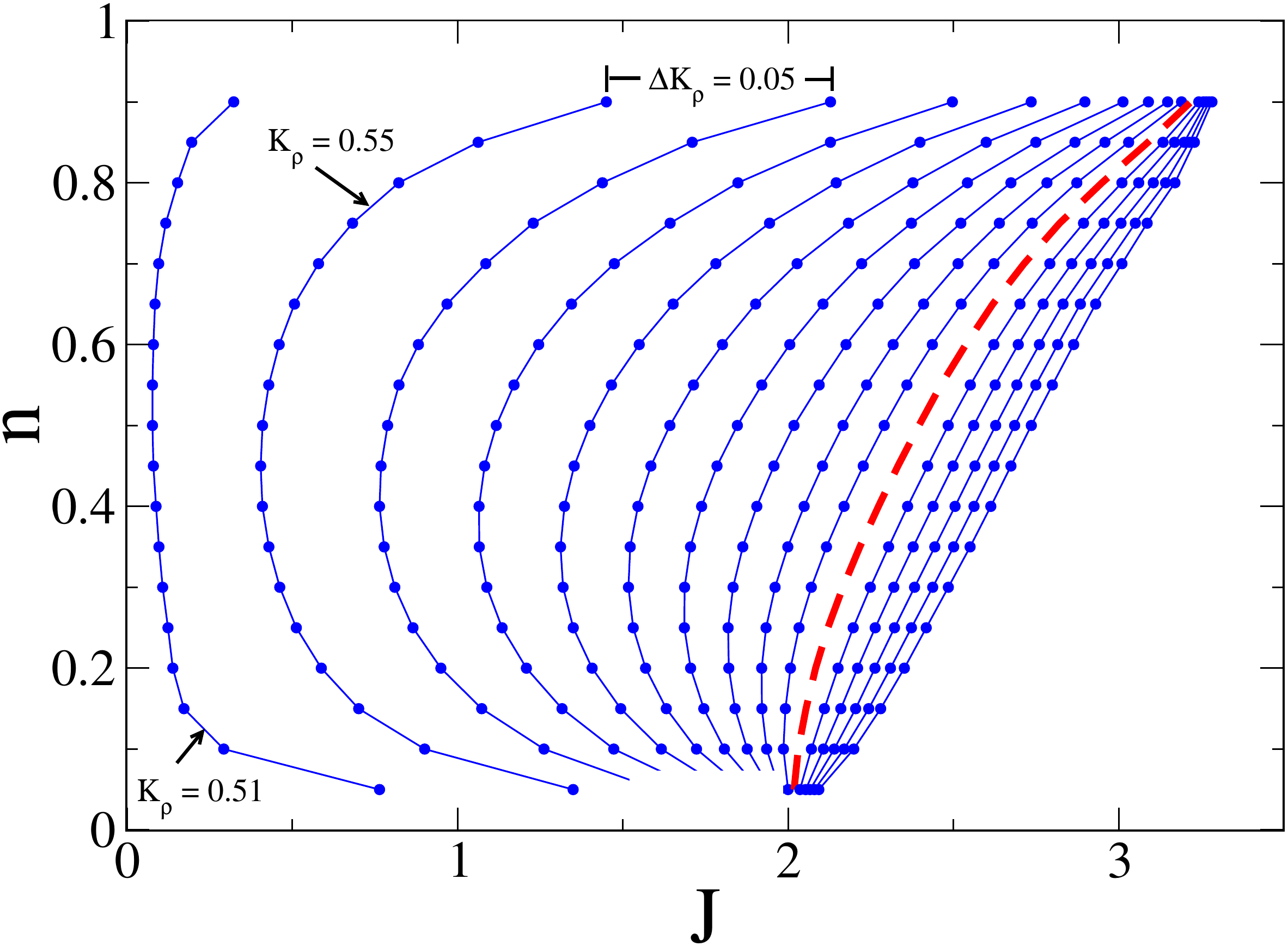}}
\caption{(color online). Regions with different Luttinger parameters $K_\rho(n,J)$. Each curve 
corresponds to a
constant value of $K_\rho$. The red (dashed) line ($K_\rho(n,J) = 1$)  denotes the boundary between the metallic phase and the superconducting phases. Note that the density of lines increases with $J$ showing the fast growth of $K_\rho$.}
\label{fig5}
\end{figure}

From the data set presented in Fig.\ \ref{fig4} we can extract all the points which fulfill $K_\rho(n,J) = const$. These are curves which separate regions with different Luttinger parameters. The different regions and curves are plotted in Fig.\ \ref{fig5}. The red (dashed) line ($K_\rho(n,J) = 1$) denotes the boundary between the metallic phase and the superconducting phases. This line and few others were also plotted in the phase diagram (Fig.\ \ref{fig1}). Note that the density of lines increases with $J$ showing the fast growth of $K_\rho$. 

\subsection{Singlet-superconductivity and spin-gap phase}
\label{spin-gap}
\begin{figure}[th!]\relax
\centerline{\includegraphics[width=3in]{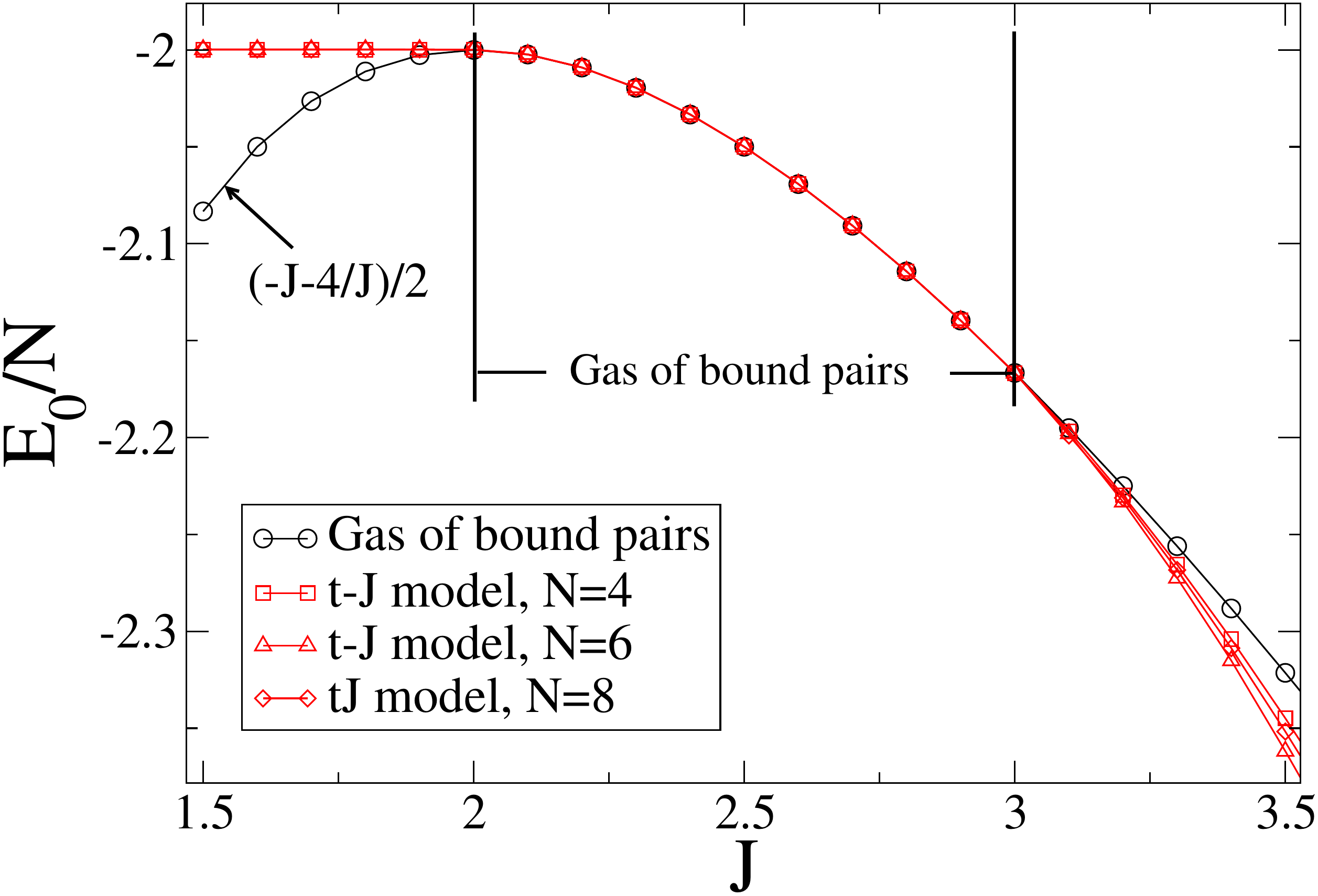}}
\caption{(color online). Energy comparison between systems containing four, six and eight particles and a gas of singlet bound pairs.  We observe a region $2.0 < J < 3.0$ where all energies are the same within an error of $10^{-5}$. This opens the possibility of the formation of a gas of singlet bound pairs and consequently of the existence of a spin gap at very low densities.}
\label{gas_bound_ene}
\end{figure}

The possibility of a region with a spin gap was first analyzed by Ogata {\em et al}.\ \cite{ogata91} for the low density limit, where a gas of singlet bound pairs may form. They compared the ground state energy of a system containing four particles to twice the energy of a system with only two particles. The energy for the last situation is $2(-J-4/J)$, where the expression $-J-4/J$ is obtained by solving exactly the Hamiltonian (\ref{Hamiltonian}) for two particles. 
We compare here the ground-state energy per particle for 4, 6, and 8 particles obtained numerically with DMRG to that of a gas of bound pairs (Fig.\ \ref{gas_bound_ene}),
using $1000$ lattice sites.
We observe a region, $2.0 < J < 3.0$, where 
energies are the same within an error of $10^{-5}$. At least from energy considerations, this is an indication of the possibility of the formation of a gas of singlet bound pairs and consequently of the existence of a spin gap at very low densities. Moreover, no evidence for the formation of more complex entities can be seen.

\begin{figure}[th!]\relax
\centerline{\includegraphics[width=3in]{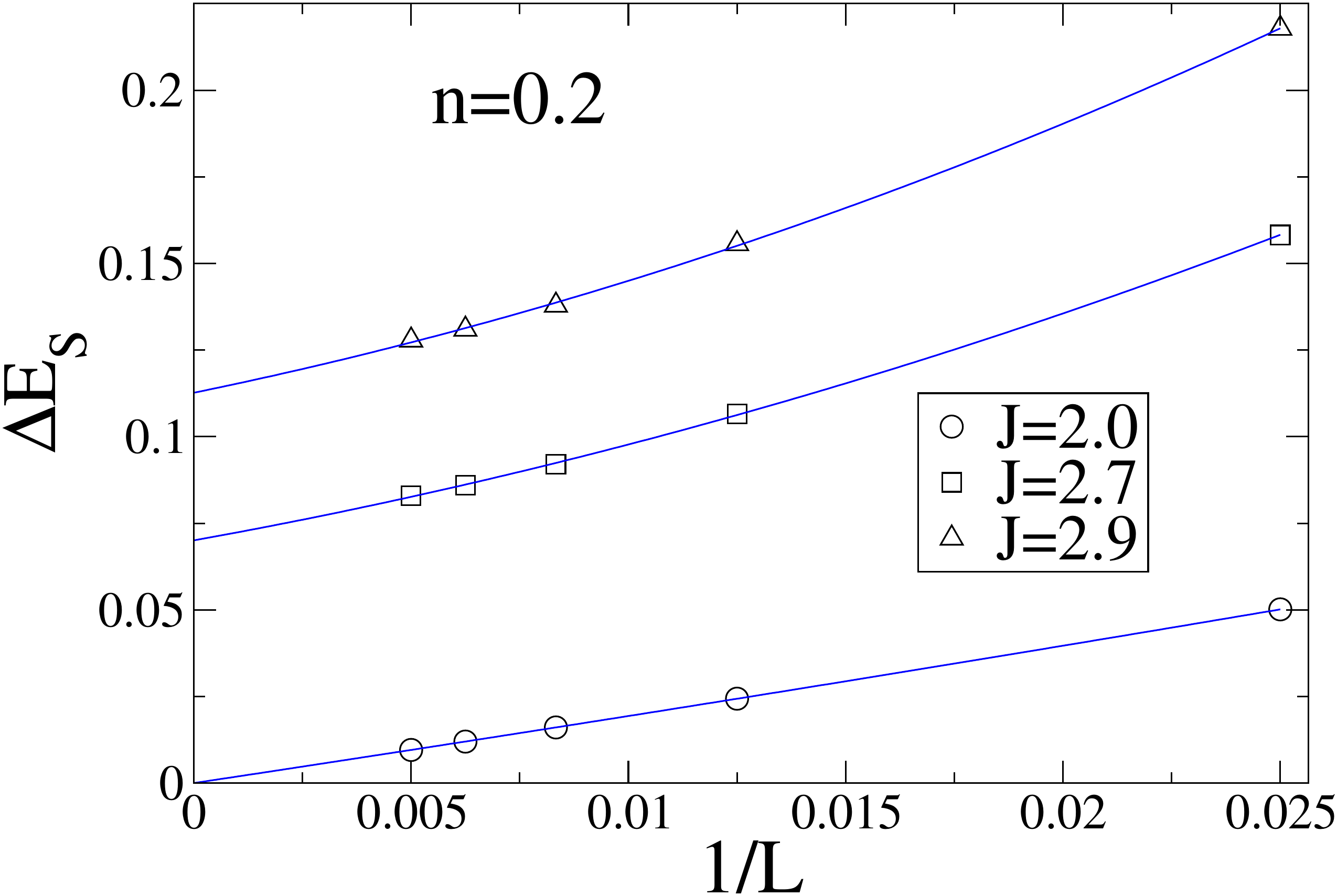}}
\caption{(color online). $\Delta E_S$ vs $1/L$ for $n=0.2$ and various values of $J$. For $J=2$, where the system is still in the metallic phase, the spin gap extrapolates to zero in the thermodynamic limit. }
\label{fig6}
\end{figure}

A more precise estimate can be obtained by measuring directly the spin gap $\Delta E_S$. 
The spin excitation energy from a singlet to a triplet state is given by the energy difference
\begin{equation}
\Delta E_S = E_0 (N, S_{tot}^z=1) -  E_0 (N, S_{tot}^z=0),
\label{e7}
\end{equation} 
where the subindex $0$ means that we take the lowest energy level with given quantum numbers $N$ and $S_{tot}^z$. In order to go to the thermodynamic limit we plot $\Delta E_S$ as a function of $1/L$ and we extrapolate to $1/L \rightarrow 0$ using $L=40,80,120,160$, and 200. Figure \ref{fig6} shows $\Delta E_S$ vs $1/L$ at $n=0.2$ and for $J=2.0$, $2.7$ and $2.9$. The extrapolations to the thermodynamic limit are performed with third-order polynomials. While for $J=2$, where the system is still in the metallic regime 
the gap extrapolates to zero,  for $J=2.7$ and $2.9$ a finite gap can be clearly seen.

\begin{figure}[th!]\relax
\centerline{\includegraphics[width=3in]{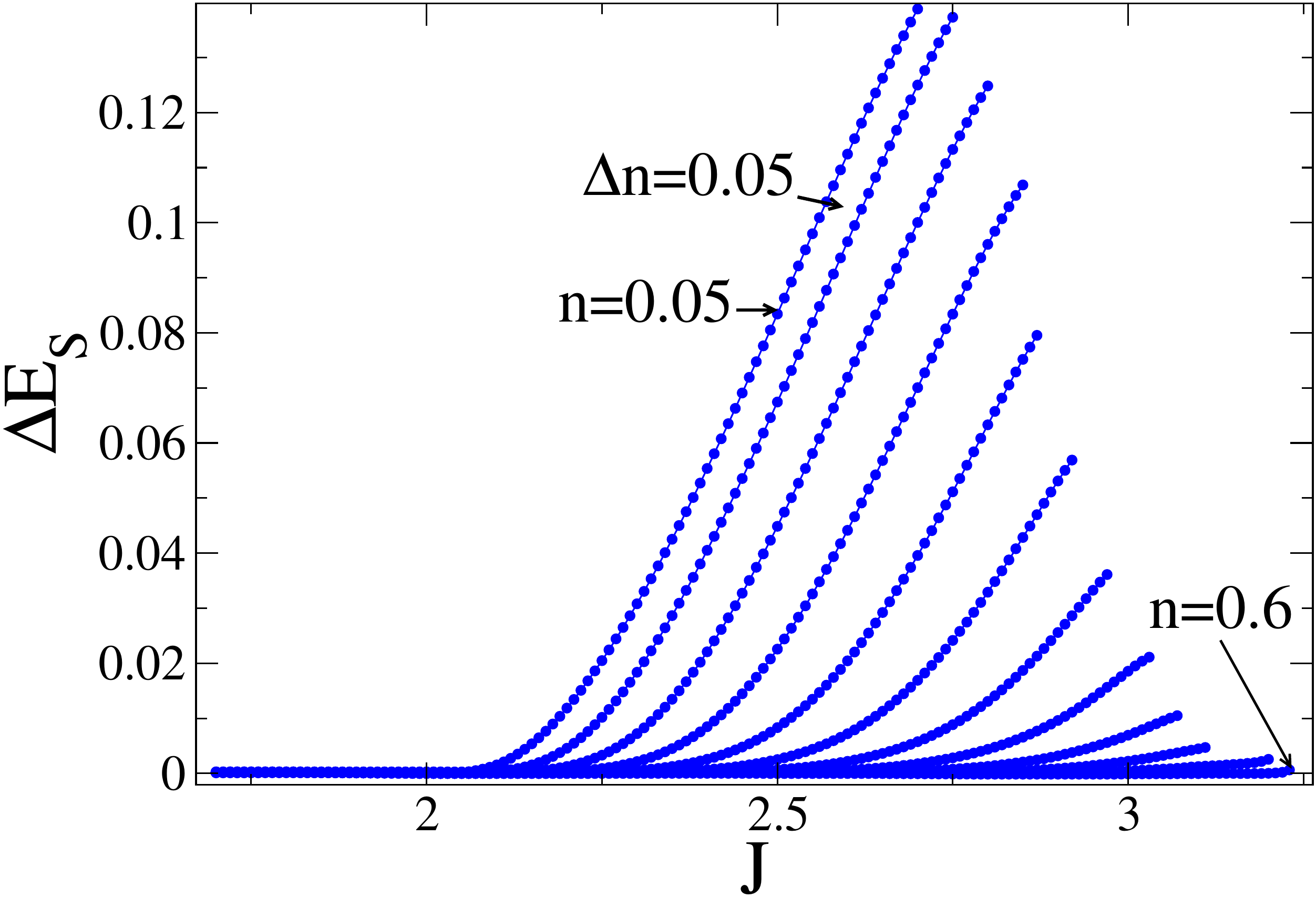}}
\caption{(color online). Spin gap $\Delta E_S$ in the thermodynamic limit as a function of $J$ and for different densities $n$. For $J < 2$ (metallic phase) the spin gap is zero for all densities. For $J > 2$ a finite spin gap emerges with a value that increases for diminishing densities. We can observe a small but finite spin gap up to $n=0.55$.}
\label{fig8}
\end{figure}

Proceeding in the same manner for different values of $n$ and $J$, we can obtain the spin gap in the thermodynamic limit (Fig.\ \ref{fig8}). For $J < 2$ (metallic phase) we observe that the spin gap is zero for all densities. For $J > 2$ a finite spin gap emerges that increases as the density diminishes. For definiteness,  
$J_c$ is defined in our case as the value of $J$ for which $\Delta E_S > 10^{-4} t$, this value being the  
the range on which $\Delta E_S$ fluctuates around zero before it definitively increases as a function of $J$ for a given density. In this manner we have obtained the lowest boundary of the spin-gap phase in the phase diagram (Fig.\ \ref{fig1}). 

\begin{figure}[th!]\relax
\centerline{\includegraphics[width=3in]{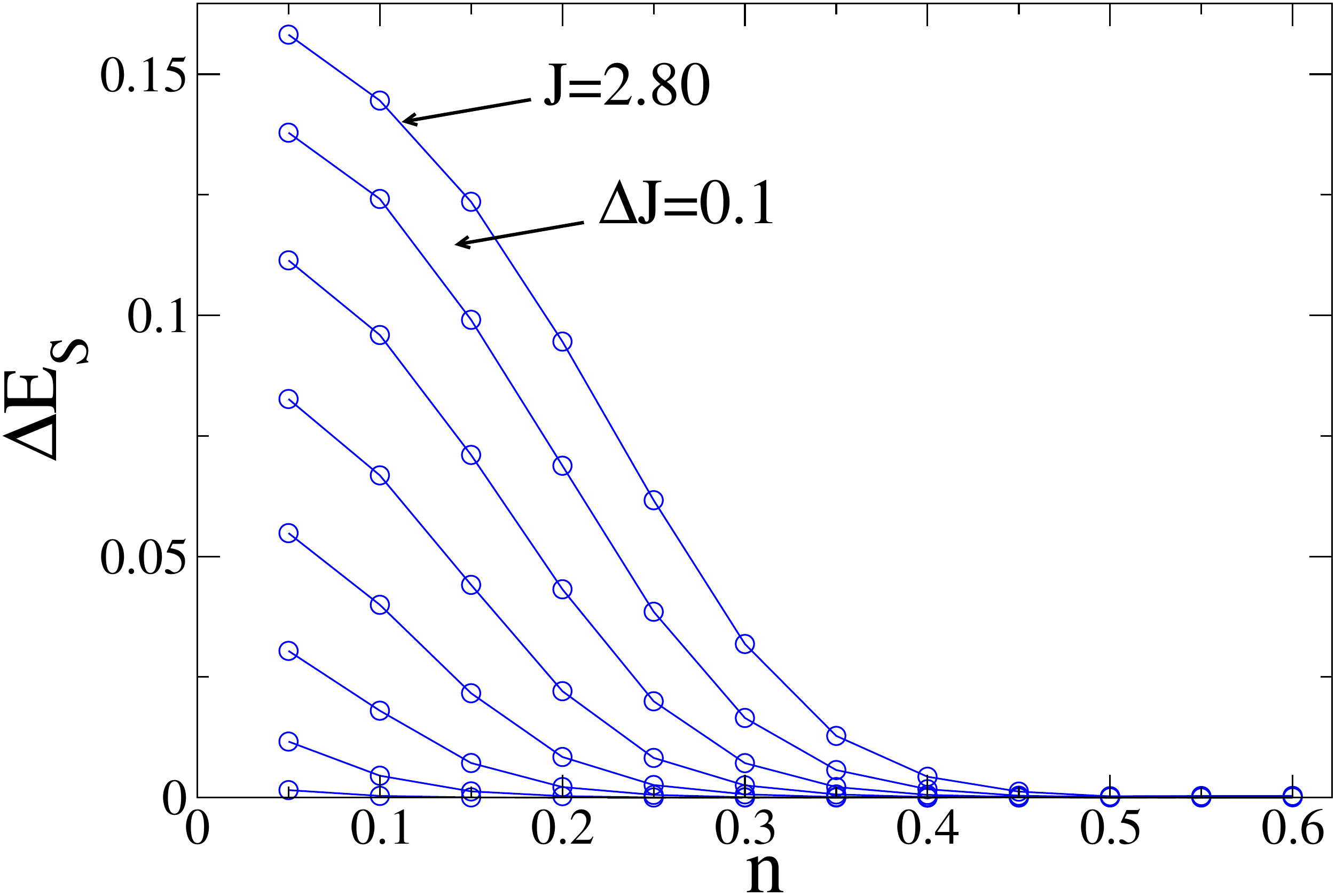}}
\caption{(color online). Spin gap $\Delta E_S$ in the thermodynamic limit as a function of $n$ for $J=2.1-2.8$.}
\label{gap_vs_rho}
\end{figure}

We present also in Fig.\ \ref{gap_vs_rho} the spin gap as a function of $n$ for $J=2.1-2.8$. Note that $\Delta E_S$ smoothly closes to zero when the density is increased. $\Delta E_S$ attains its largest values as $J$
increases for the limit of vanishing densities, reaching $\Delta E_S \approx J/20$. Since in one dimension superconductivity can only set in at temperature $T=0$, such a finite value of the gap gives the energy scale at which pairs form, signaling the existence of preformed pairs in this regime.  A further increase of $J$ leads to phase separation, that we discuss next. 
A detailed analysis of the transition to phase separation, centering on the possibility of clusters with more than one pair is presented in Sec.\ \ref{PairingRS}.

\subsection{Phase separation}
In this phase the attraction among the particles is so strong that they start to form antiferromagnetic domains, such that the system separates into particle- and hole rich regions. In the limit $J \rightarrow \infty$ all the particles join in a single island, which can be described by the Heisenberg model forming an electron solid phase, a denomination proposed by Chen and Moukouri\cite{chen96},
where the kinetic fluctuations are strongly quenched and only spin fluctuations remain. We first consider the inverse of the compressibility that vanishes at the onset of phase-separation.

\begin{figure}[th!]\relax
\centerline{\includegraphics[width=3in]{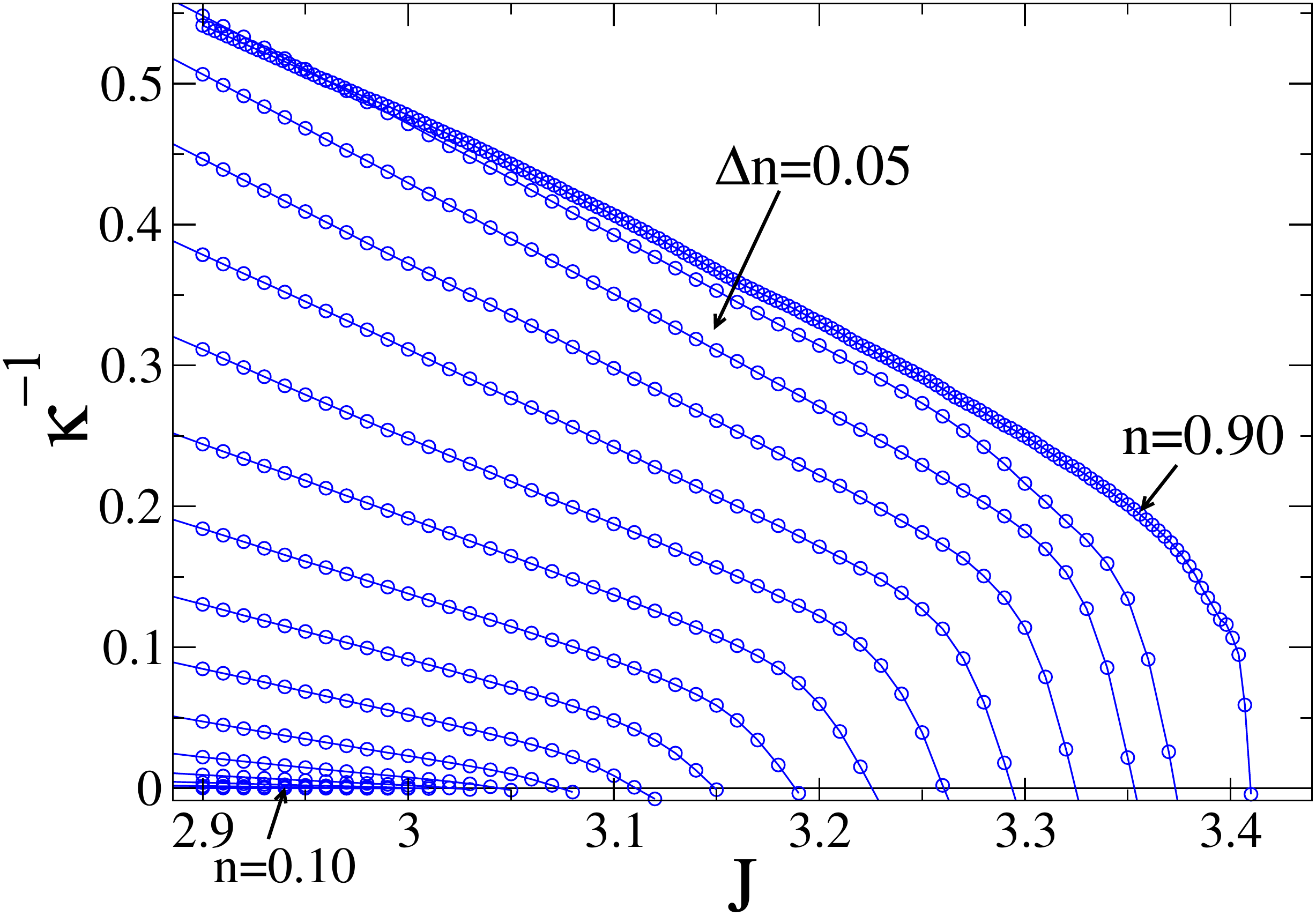}}
\caption{(color online). Inverse of the compressibility $\kappa ^{-1}$ as a function of $J$ for $n=0.1-0.9$ with $\Delta n = 0.05$.}
\label{fig9}
\end{figure}
At zero temperature the expression for the inverse compressibility is given by
\begin{eqnarray}
\kappa ^{-1} (n)  & = & n^2 \frac{\partial ^2 e_0(n)}{\partial n^2}
\nonumber \\ & \approx & 
 n^2 \frac{[e(n+\Delta n) + e(n-\Delta n) - 2e(n)]}{\Delta n^2},
\label{e9}
\end{eqnarray}
where $e_0(n) = E_0/L$ is the energy density per site, and the second line gives the approximation for finite
($\Delta n = 0.05$) changes in the density. 

\begin{figure}[th!]\relax
\centerline{\includegraphics[width=3in]{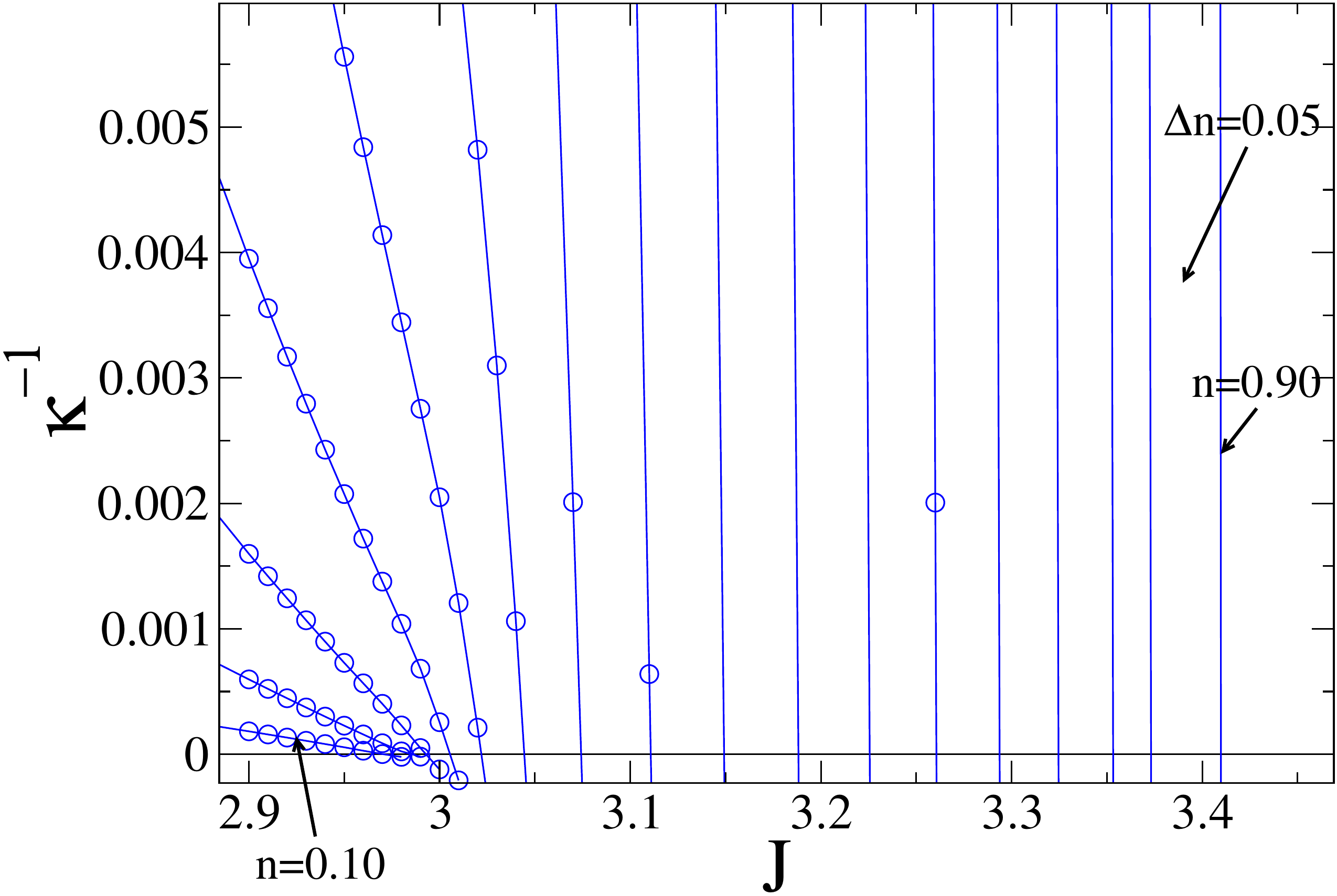}}
\caption{(color online). Zoom of Fig.\ \ref{fig9}. The points $(n,J_c)$ where $\kappa ^{-1} = 0$ define the boundary of the phase-separated phase (infinite compressibility).}
\label{fig10}
\end{figure}
For the extrapolation of $e(n)$ we use $L=40,80,120,160$ and a third-order polynomial fitting. Figure \ref{fig9} shows $\kappa ^{-1}$ vs $J$ for different densities. At low densities $\kappa ^{-1}$ is vanishingly small, making the extraction of $J_c$ very difficult in that region. In order to see more clearly the critical value $J_c(n)$ where $\kappa ^{-1}$ vanishes, we display a zoom of Fig.\ \ref{fig9} 
in Fig.\ \ref{fig10}. In this manner we found the boundary of the phase-separated phase in Fig.\ \ref{fig1}.  Note that, in comparison to other studies \cite{ogata91,chen93,helberg93}, the phase separation boundary is shifted to higher values of $J$.

\section{Correlation functions and density in real space}

\subsection{Structure factors}
\label{struc-fact}

\begin{figure}[th!]
\begin{center}
$\begin{array}{c}
\includegraphics[width=3in]{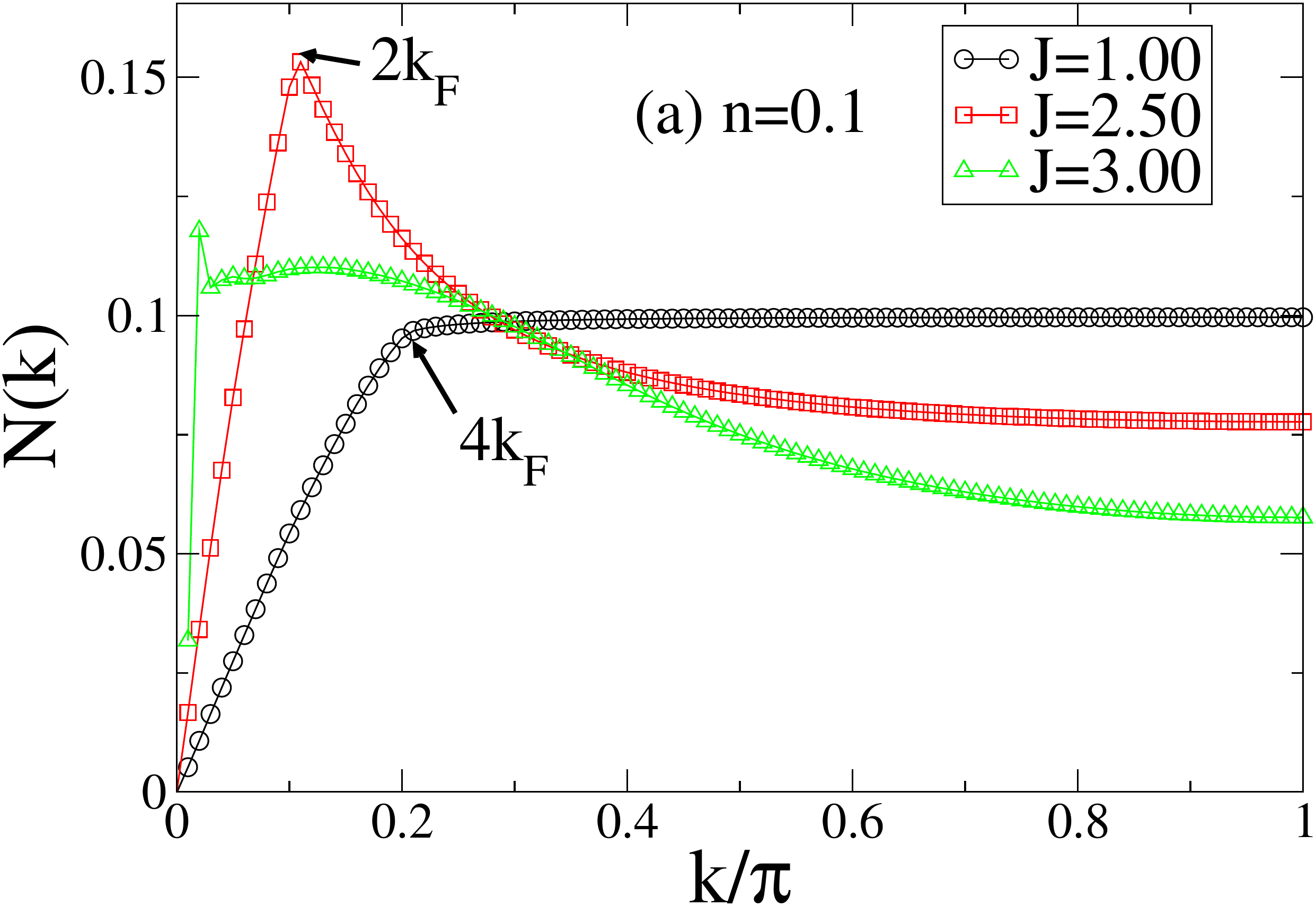}  \\ 
\includegraphics[width=3in]{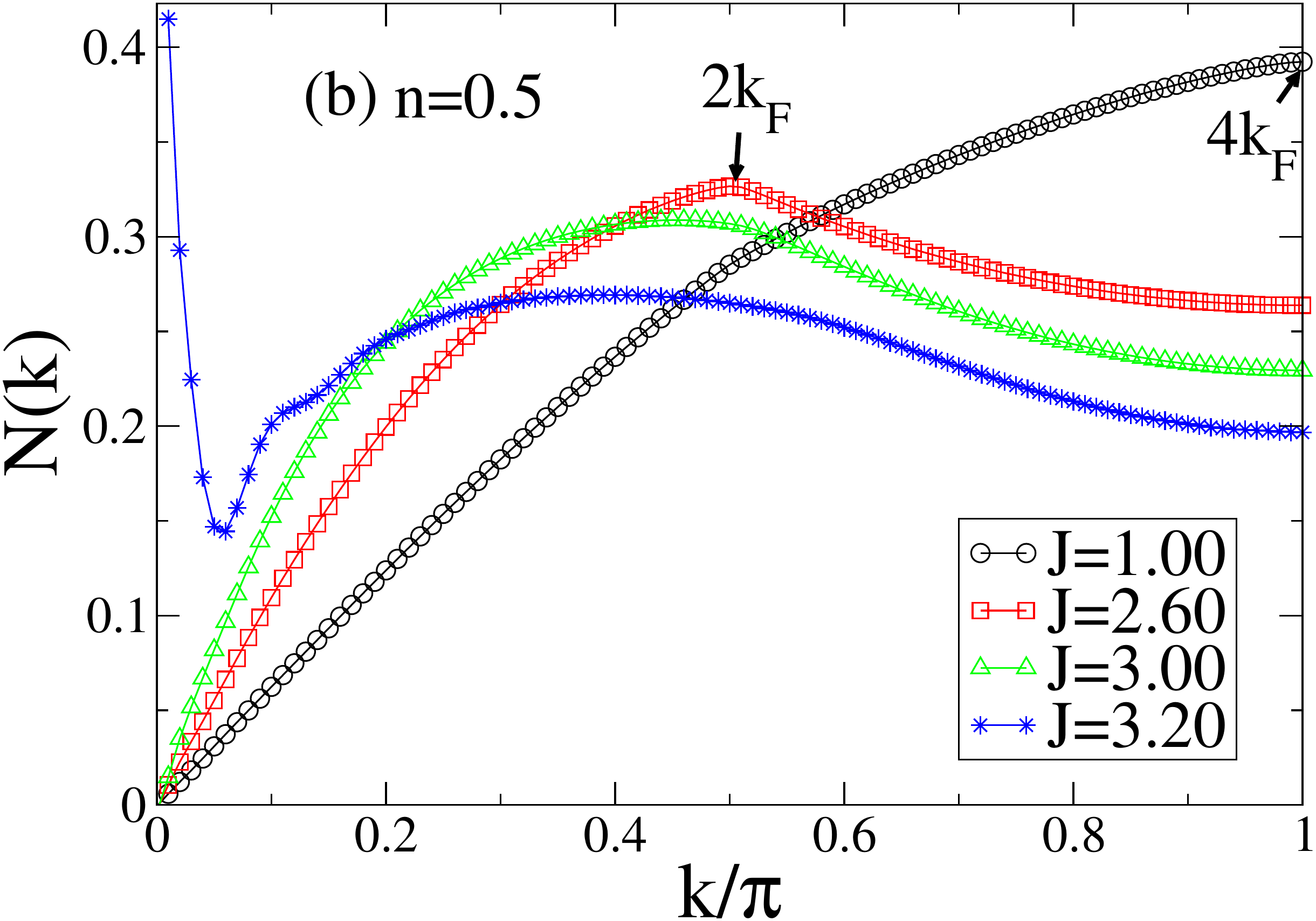} \\
\includegraphics[width=3in]{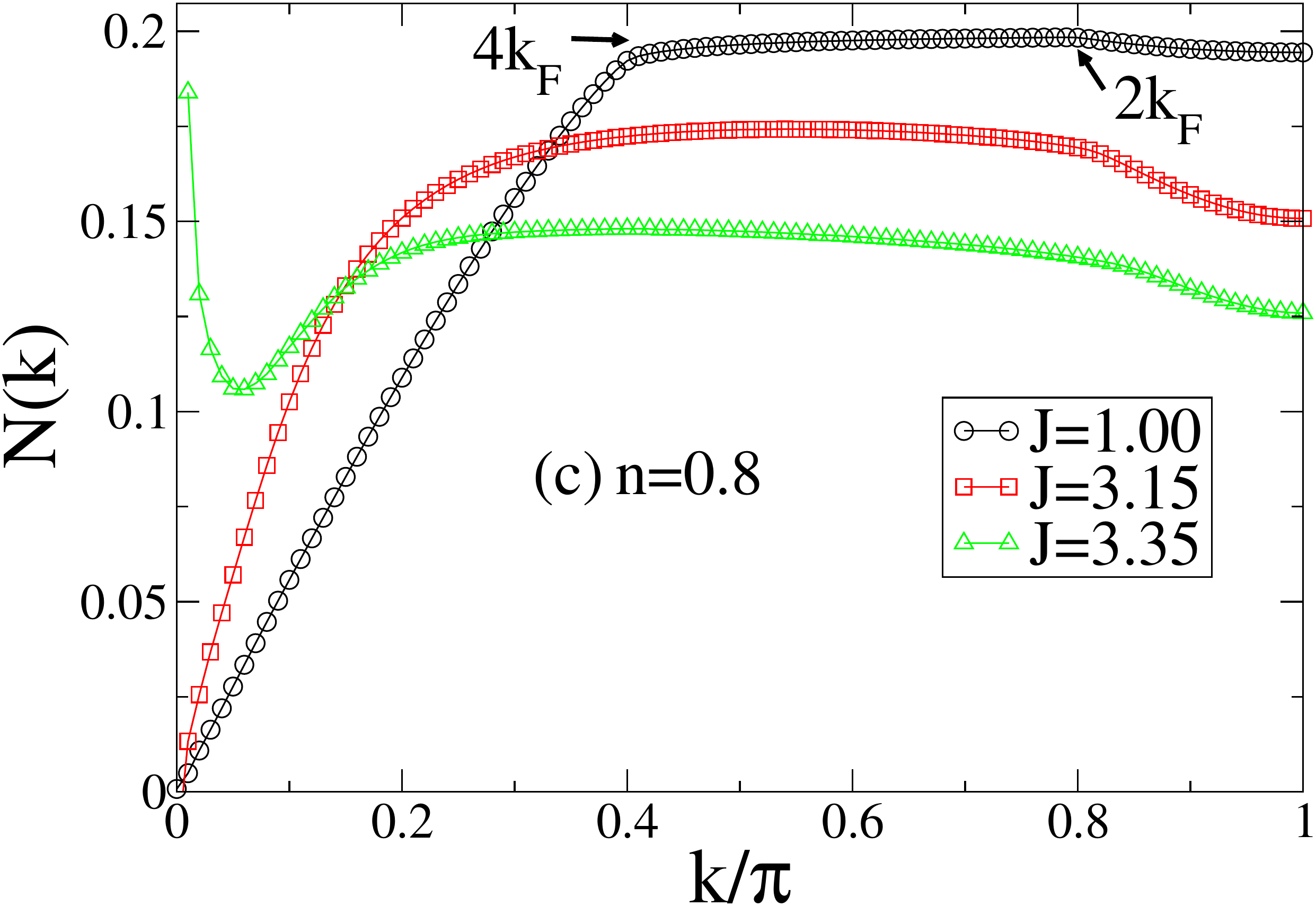}  \\ 
\end{array}$
\end{center}
\caption{(color online). Structure factor $N(k)$ for the density-density correlation function for $L=200$ and for different values of $n$ and $J$. 
The location of $2 k_F$ and $4 k_F$ correspond to the ones resulting from folding them back to the sector $k>0$ in the first Brillouin zone.}
\label{Nk}
\end{figure}

\begin{figure}[ht!]
\begin{center}
$\begin{array}{c}
\includegraphics[width=2.9in]{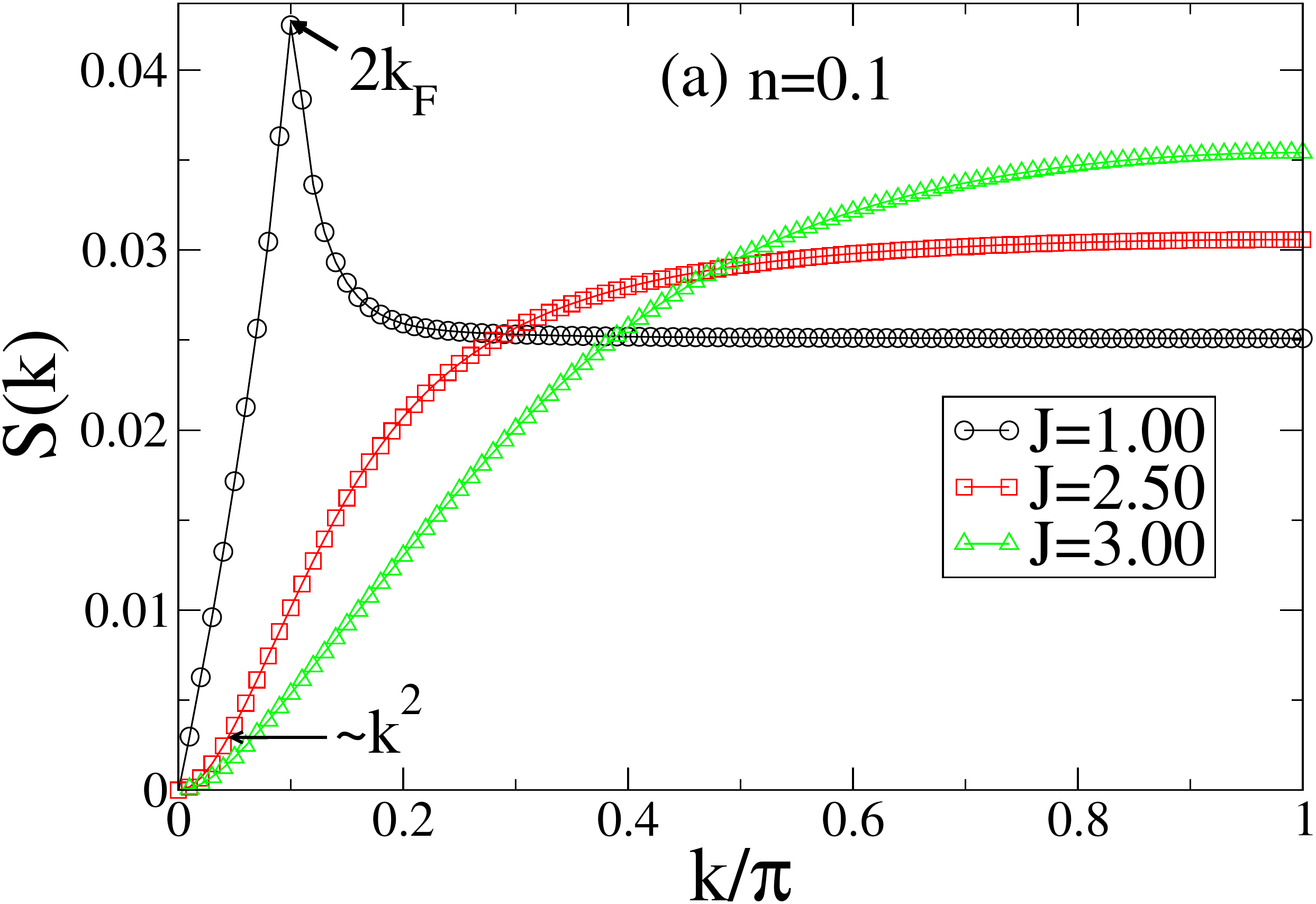}  \\ 
\includegraphics[width=2.9in]{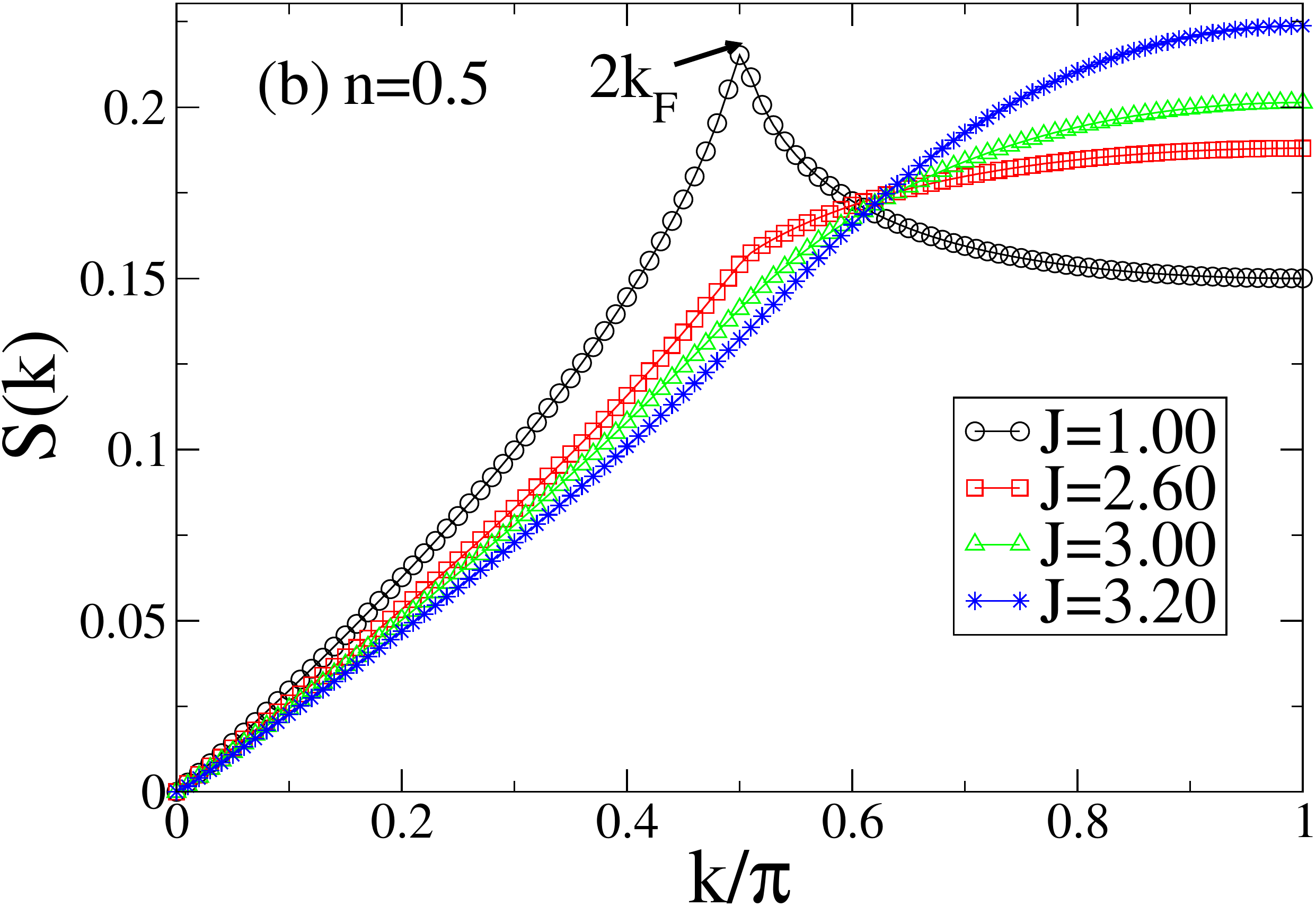} \\
\includegraphics[width=2.9in]{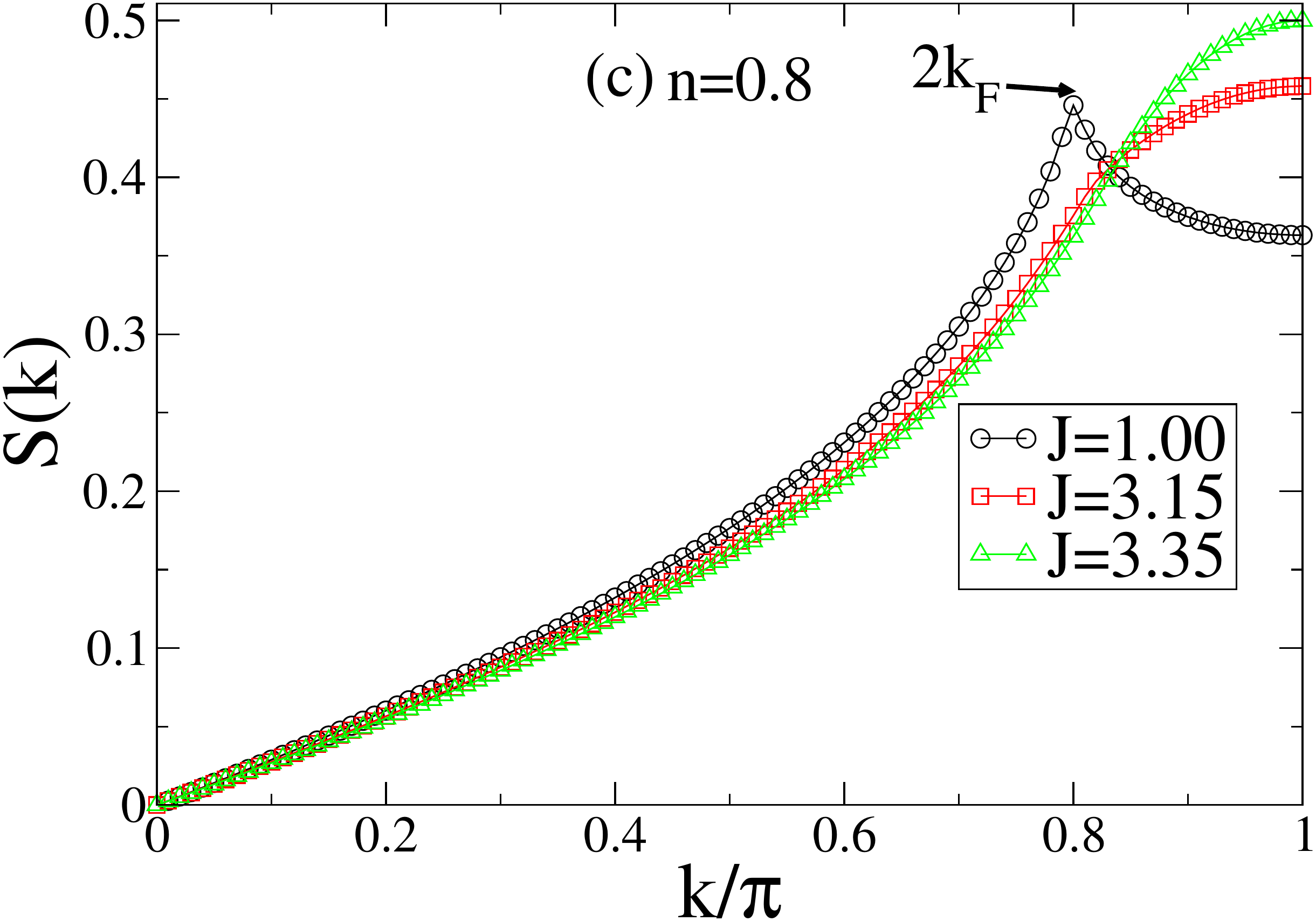}  \\ 
\end{array}$
\end{center}
\caption{(color online). Structure factor $S(k)$ for the
spin-spin correlation function  for $L=200$ and for different values of $n$ and $J$.}
\label{Sk}
\end{figure}

\begin{figure}[hbt!]
\begin{center}
$\begin{array}{c}
\includegraphics[width=3in]{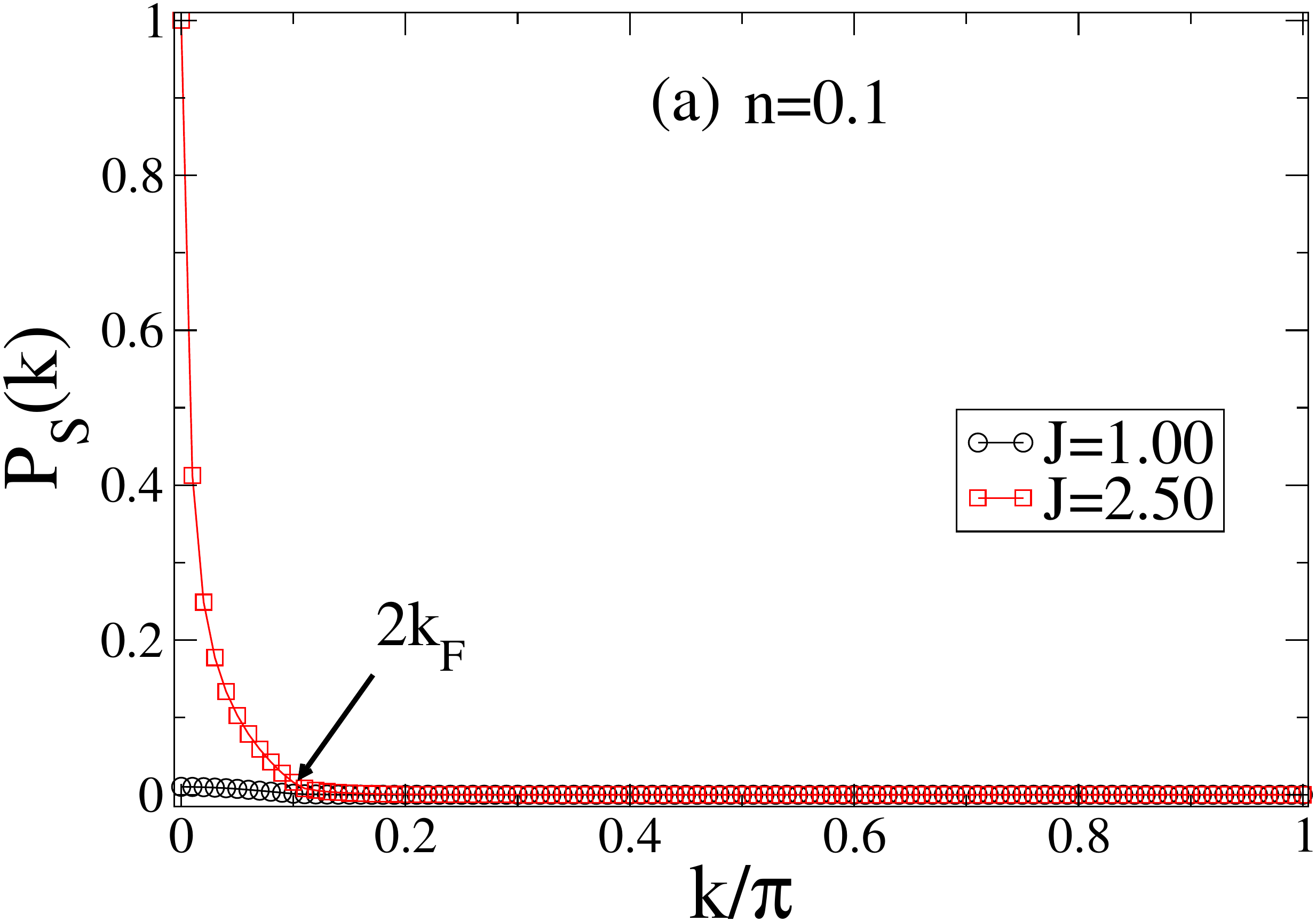}  \\ 
\includegraphics[width=3in]{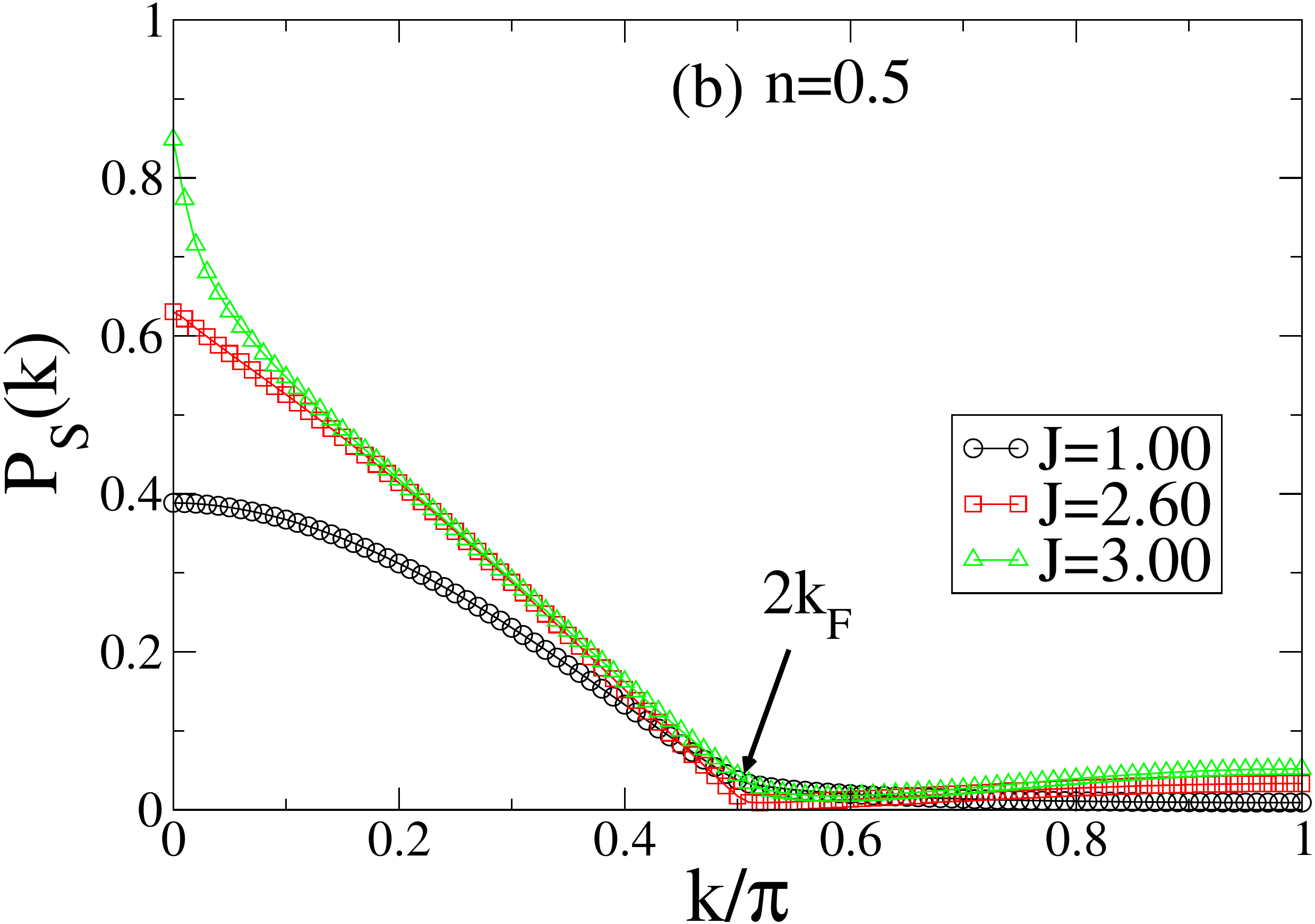} \\
\includegraphics[width=3in]{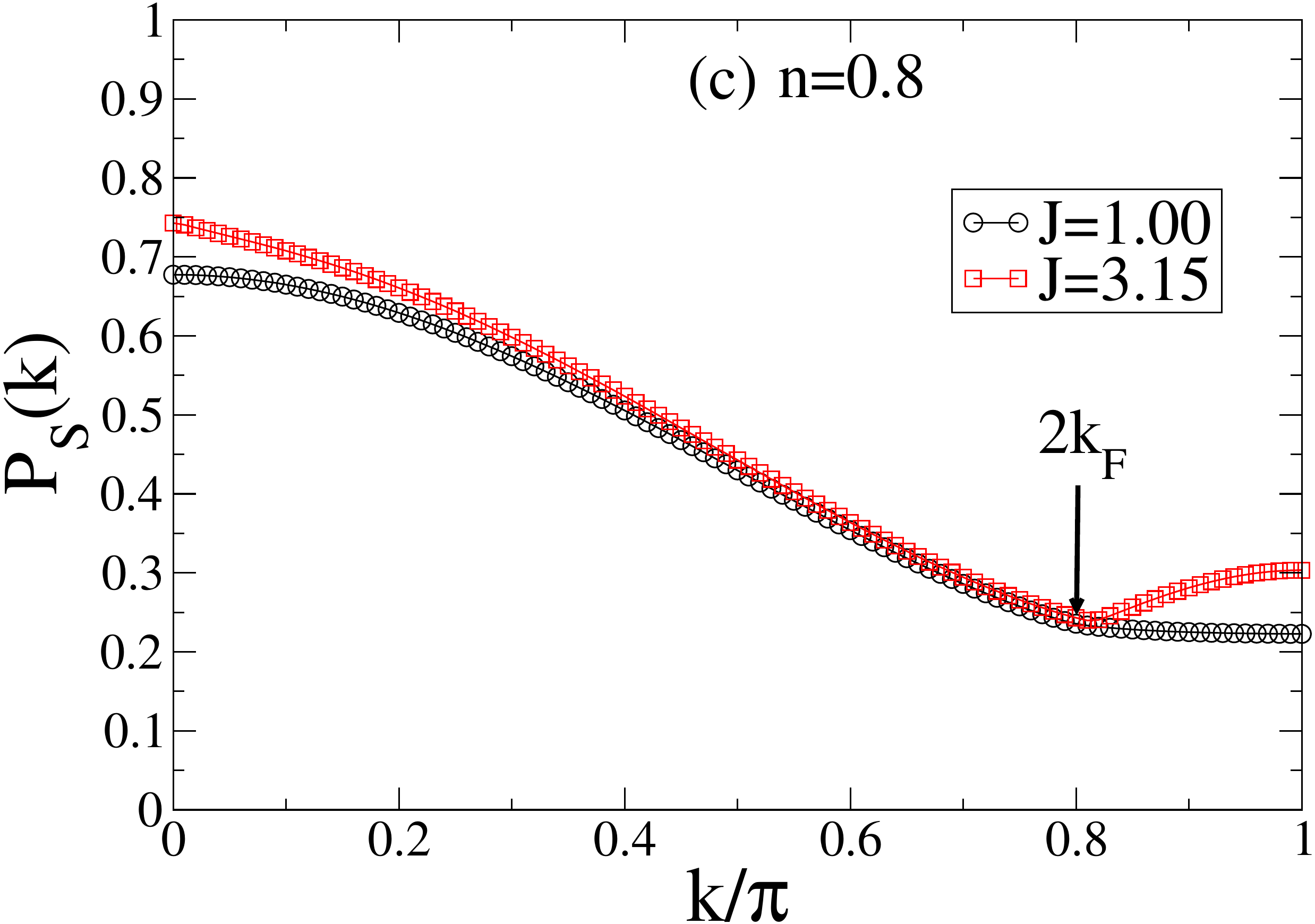}  \\ 
\end{array}$
\end{center}
\caption{(color online). Structure factor $P_S(k)$ for the singlet pair-pair correlation function for $L=200$ and for different values of $n$ and $J$.}
\label{Pk}
\end{figure}

\begin{figure}[ht!]
\begin{center}
$\begin{array}{c}
\includegraphics[width=3in]{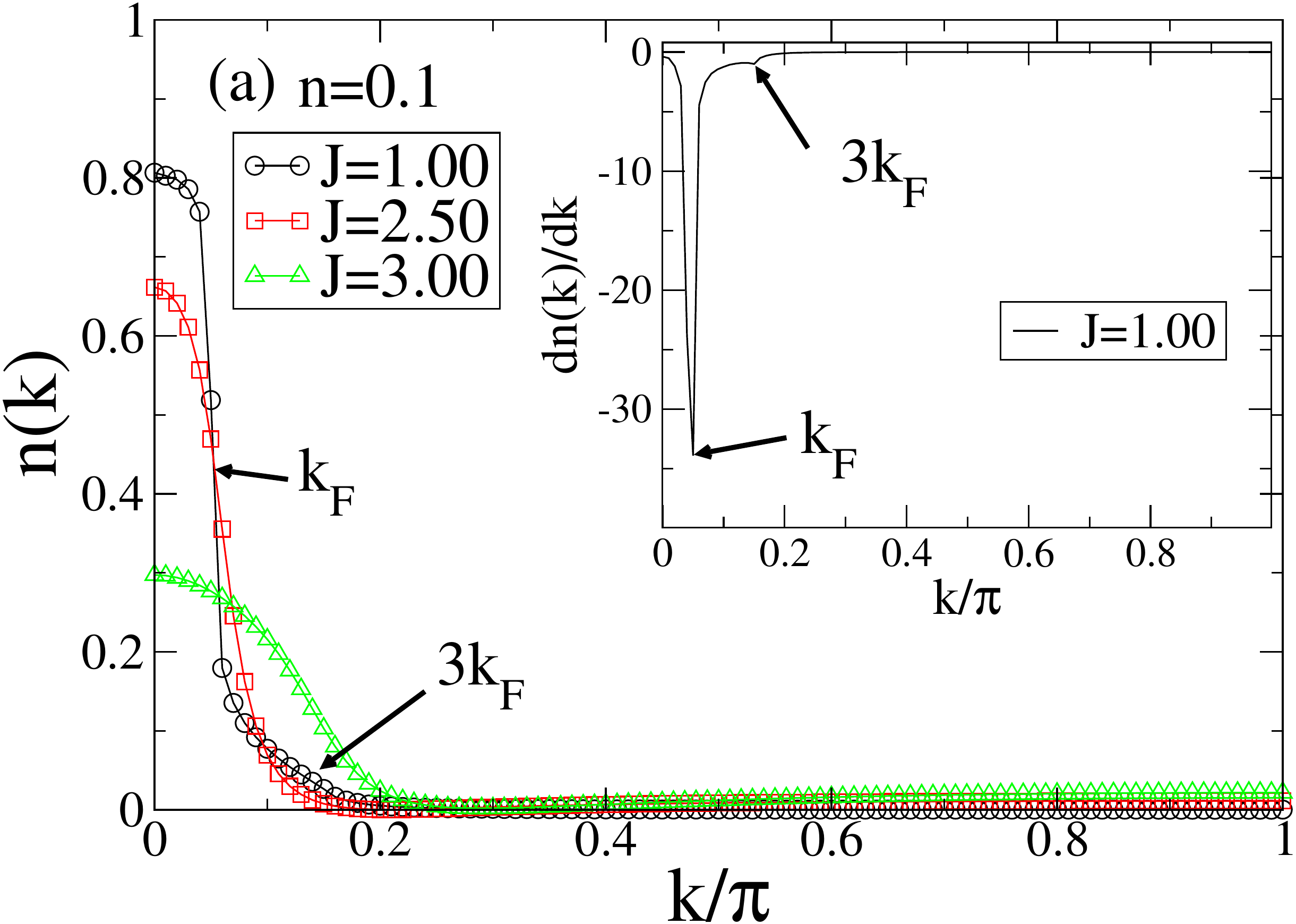}  \\ 
\includegraphics[width=3in]{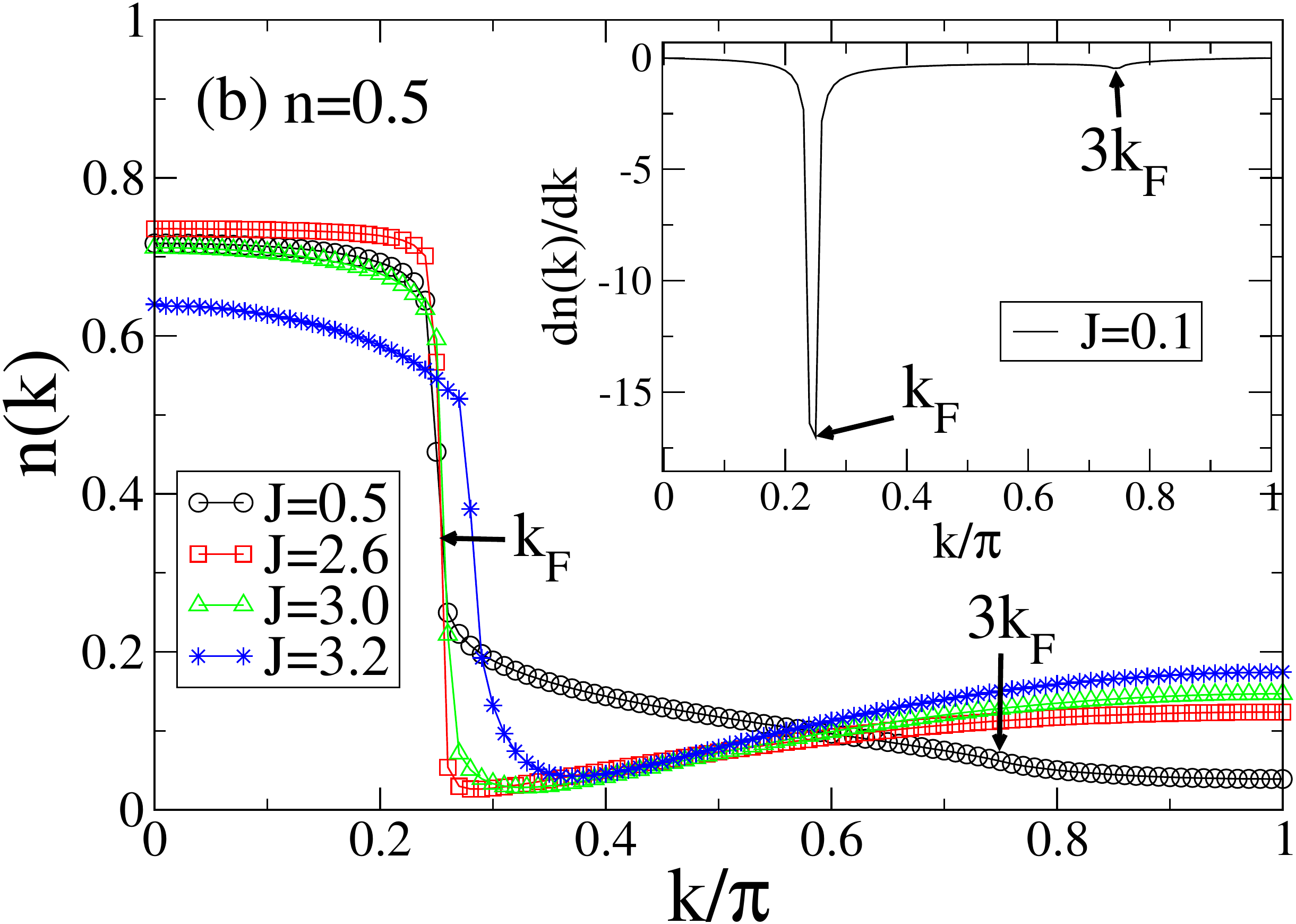} \\
\includegraphics[width=3in]{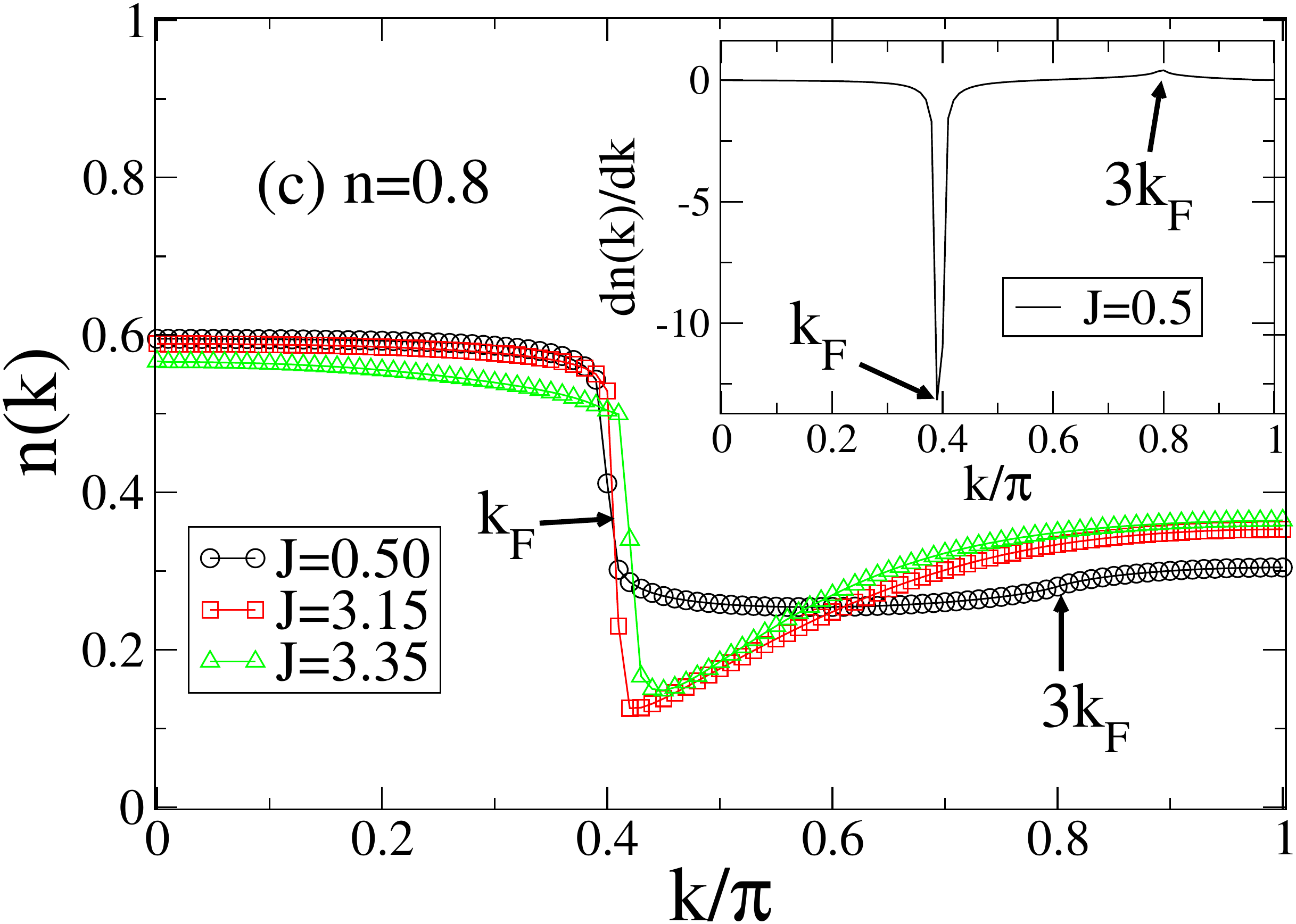}  \\ 
\end{array}$
\end{center}
\caption{(color online). Momentum distribution function $n(k)$ for $L=200$ and for different values of $n$ and $J$.}
\label{nk}
\end{figure}

In order to provide a more detailed characterization of the different phases in the phase diagram determined in the previous section, we consider here the structure factors Eqs.\ (\ref{e1}) - (\ref{e5}). 
They are shown in Figs.\ \ref{Nk}-\ref{nk} for $L=200$ and for different values of $n$ and $J$. 

Figure \ref{Nk} displays the structure factor $N(k)$ for the density-density correlation function, that for a LL is as follows \cite{schulz90}:
\begin{eqnarray}
\label{ncorr}
\langle n(r)n(0) \rangle  & = & \frac{K_\rho}{(\pi  r)^2} 
+ A_1 \frac{\cos(2k_F r)}{r^{1+K_\rho}} \ln^{-3/2} (r)
\nonumber \\ &  & 
+ A_2 \cos(4k_F r) r^{-4K_\rho} \; ,
\end{eqnarray}
where $A_1$ and $A_2$ depend in general on the parameters of the model.
For each density we have chosen values of $J$ within one of the phases. For $J=1$, a LL phase is realized for all densities, such that a $4k_F$ ($k_F=n\pi /2$) anomaly typical for a repulsive LL can be observed. 
As shown by Eq.\ (\ref{ncorr}), it is strongest for the smallest values of $K_\rho$, that are essentially achieved for all densities for $J \leq 1$, as shown by Fig.\ \ref{fig4}.
As $J$ increases, the $4k_F$ anomaly is suppressed, and a $2k_F$ cusp is formed, signaling a 
$2k_F$ charge density wave due to the tendency towards pairing that is enhanced by $J$ \cite{giamarchi04, assaad91}. However, going closer to the boundary to phase separation (see the curves for $J=3$ in Fig.\ \ref{Nk} (b), and for $J=3.15$ in Fig.\ \ref{Nk} (c)), the singularity is rounded, and increasing $J$ even more, such that phase-separation is reached, leads to the development of a singularity in $N(k)$ around $k=0$. 
That is, the system starts to develop an instability towards long-wavelength charge fluctuations signaling the appearance of phase-separation. However, since in the finite system simulations, $N(k=0)=0$ for all cases due to the conservation of total particle number, such a singularity can only be followed up to the smallest non-vanishing value of momentum for a given system size. 
Except for the cases, where the system enters the phase separated phase, $N(k)$ goes continuously to zero, as $k \rightarrow 0$, such that $K_\rho$ can be extracted, as discussed in Sec.\ \ref{Sec:PhaseDiagram}. In both superconducting phases, $K_\rho > 1$, as expected, and increases as one goes deeper into the spin-gap phase. 

In Fig.\ \ref{Sk} the structure factor $S(k)$ for the spin-spin correlation function, is shown. The spin-spin correlation function is given by \cite{schulz90}
\begin{eqnarray}
\label{Scorr} 
\langle S^Z (r) S^Z (0) \rangle  =  \frac{1}{(\pi  r)^2} 
+ B_1 \frac{\cos(2k_F r)}{r^{1+K_\rho}} \ln^{1/2} (r).
\end{eqnarray}
The tendency to antiferromagnetism in the t-J model 
can be observed at $J = 1$ for all densities, as revealed by a sharp cusp at $2k_F$ in $S(k)$, corresponding to quasi-long range order in the magnetic channel. 
However, on increasing $J$, 
such that the system enters the superconducting phase, the sharp peak is suppressed. For densities below $n=0.6$, a spin gap develops, such that on entering this Luther-Emery (LE) phase, due to the exponential decay of the spin-spin correlation function, the singularity at $2k_F$ is completely suppressed \cite{giamarchi04}.
In this case, the corresponding structure factor has a quadratic behavior at small $k$'s \cite{chen93}. This fact can be clearly observed only at $J = 2.5$ and $n=0.1$, where the spin gap is well developed. As shown in Fig.\ \ref{fig8}, the spin gap at $J=3$ for a density $n= 0.5$ is extremely small ($\Delta E_S \sim 10^{-3}$), such that extremely large systems would be required to show such a behavior. 
On increasing $J$ a maximum at $k=\pi$ develops due to the tendency to form antiferromagnetic islands in the phase-separated region. However, while the charge structure factor already shows clear signals of phase-separation (Fig. \ref{Nk}), $S(k)$ displays only a broad maximum around
$k=\pi$ not yet indicative of antiferromagnetic quasi-long range order. 
A larger value of $J$ is necessary in order to achieve such a state, as suggested by the curves corresponding to the largest value of $J$ in Fig.\ \ref{Sk}. 
Hence,  the onset of phase-separation and the formation of an antiferromagnetic island do not occur simultaneously, as already observed in an earlier quantum Monte Carlo study \cite{assaad91}. 
We will discuss the formation of antiferromagnetic islands in more detail in Sec.\ \ref{PairingRS}.

In Fig.\ \ref{Pk} the structure factor $P_S (k)$ for the singlet pair-pair correlation function is shown. The corresponding correlation function in the LL sector is given by \cite{giamarchi04}
\begin{eqnarray}
\langle  \Delta _S ^\dagger (r) \Delta _S (0) \rangle  & = & C_0 r^{-(1+1/K_\rho)} 
\nonumber \\ 
&  & + C_1 \cos(2k_F r)r^{-(K_\rho+1/K_\rho)},
\label{PScorr} 
\end{eqnarray}
and in the LE phase by
\begin{eqnarray}
\langle  \Delta _S ^\dagger (r) \Delta _S (0) \rangle  & = & C_0' r^{-1/K_\rho}
\nonumber \\ & &
+ C_1' \cos(2k_F r) r^{-(K_\rho +1/K_\rho)}
,
\label{PScorr-gap}
\end{eqnarray}
where we have ignored logarithmic corrections.
Noticing the different powers appearing in Eqs.\ (\ref{PScorr}) and (\ref{PScorr-gap}), and taking into account that $K_\rho > 1$ in the superconducting phases, it can be seen that while in the LL sector $P_S(k)$
does not present a divergence, it will have a diverging contribution at $k=0$ in the LE case. Such a behavior can be seen in Fig.\ \ref{Pk}.
Within the spin-gap phase (Fig.\ \ref{Pk} (a), $J=2.5$ and Fig.\ \ref{Pk} (b), $J=3$), a strong increase is observed for $k \rightarrow 0$, indicating the onset of quasi-long range order in this channel. 
(singlet-superconductivity). While increasing $J$ and $n$ also an enhancement of $P_S(k=\pi)$ is produced in the LL phase, the curves display only a rounded maximum in this case.  
We observe also a $2k_F$ anomaly in $P_S(k)$ that correlates with the 
$2k_F$ singularities in $N(k)$, displayed in Figs.\ \ref{Nk}. We will discuss these correlation functions further in Secs.\ \ref{SCgapless} and \ref{CDW_vs_SS}, where they will be confronted with other possible orderings like triplet pairing and charge-density wave (CDW) formation.

In Fig.\ \ref{nk} the momentum distribution function $n(k)$ is shown. In the metallic phase an edge develops in $n(k)$ that evolves into a singularity in the thermodynamic limit, and hence, defines a Fermi surface, in the way expected for a LL. A weak anomaly at $3k_F$ can be observed, as revealed by the derivative of $n(k)$ shown in the insets in Fig.\ \ref{nk}.
As was analytically shown by Ogata and Shiba \cite{ogata90} for the Hubbard model, this singularity in $n(k)$ is related to the fact that one electron close to $k_F$ can be excited to a state close to $3k_F$ together with an electron-hole pair excitation having its momentum near $-2k_F$. In Fig.\ \ref{nk} the $3k_F$ anomaly disappears as $J$ increases, in agreement with previous DMRG studies \cite{chen96}.

For $n=0.1$ and $J=2.5$, which corresponds to a point in the spin-gap phase, we observe that the Fermi surface is destroyed. For the higher densities in Figs.\ \ref{nk} (b) and (c), such a flattening of $n(k)$ around $k_F$ cannot be observed because the spin gap for such parameter values is very small. An interesting feature seen for such densities is the increase of $n(k)$ for $k > k_F$, contrary to what is expected in a weakly correlated metal.  Such a behavior is indicative of the presence of spectral weight below the Fermi energy for wavevectors $k > k_F$, as was already observed in previous quantum Monte Carlo simulations 
\cite{lavalle03}. On the contrary, for $n=0.1$ and $J=2.0$ (where $K_\rho \approx 1.0$) $n(k)$ has a shape closer to  the momentum distribution in the free case.  

\subsection{Gapless superconducting phase\label{SCgapless}}
In this section we examine closer the pairing correlation functions both for singlet as well as for triplet pairing.  The long-distance behavior of the correlation function for singlets was already given in Eq.\ (\ref{PScorr}) and for triplet pairing is as follows \cite{giamarchi04, pruschke92}:
\begin{eqnarray}
\label{PTcorr} 
\langle  \Delta _T ^\dagger (r) \Delta _T (0) \rangle  & \sim & D_0 r^{-(1+1/K_\rho)} 
\nonumber \\ 
&  & + D_1 \cos(2k_F r)r^{-(K_\rho+1/K_\rho+2)},
\label{PTcorr}
\end{eqnarray}
where the operator $\Delta _T$ corresponds to triplet superconducting pairing (see Eq.\ (\ref{e3.2})). Here we have ignored logarithmic corrections. 

In the gapless superconducting phase, since there is no gap to triplet states, and the leading power-law is the same for both singlet and triplet channels, the question may arise about
the relative strength of singlet- (SS) and triplet-superconductivity (TS), as already pointed out by Pruschke and Shiba \cite{pruschke92}.

\begin{figure}[th!]\relax
\begin{center}
$\begin{array}{c}
\centerline{\includegraphics[width=3in]{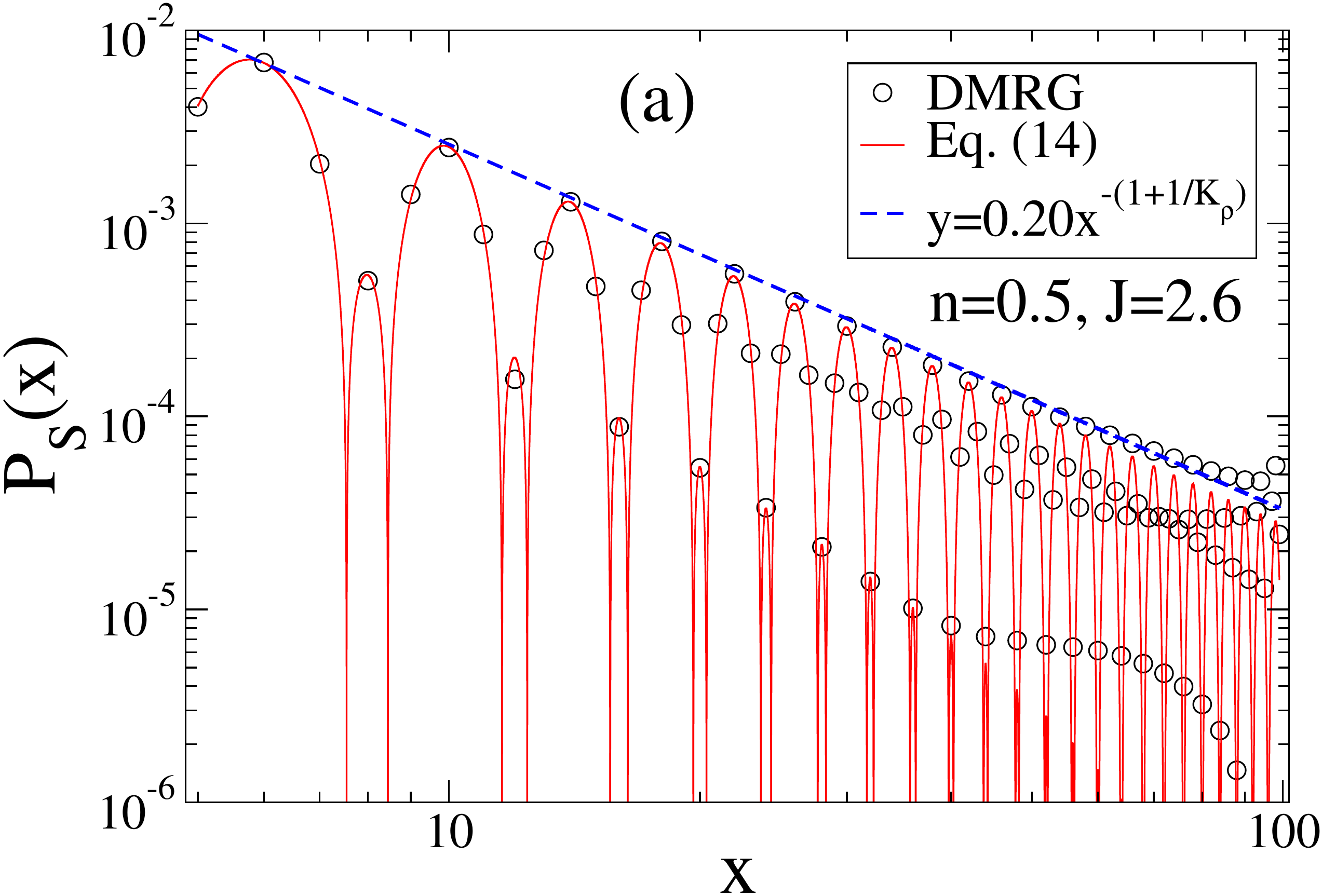}} \\
\centerline{\includegraphics[width=3in]{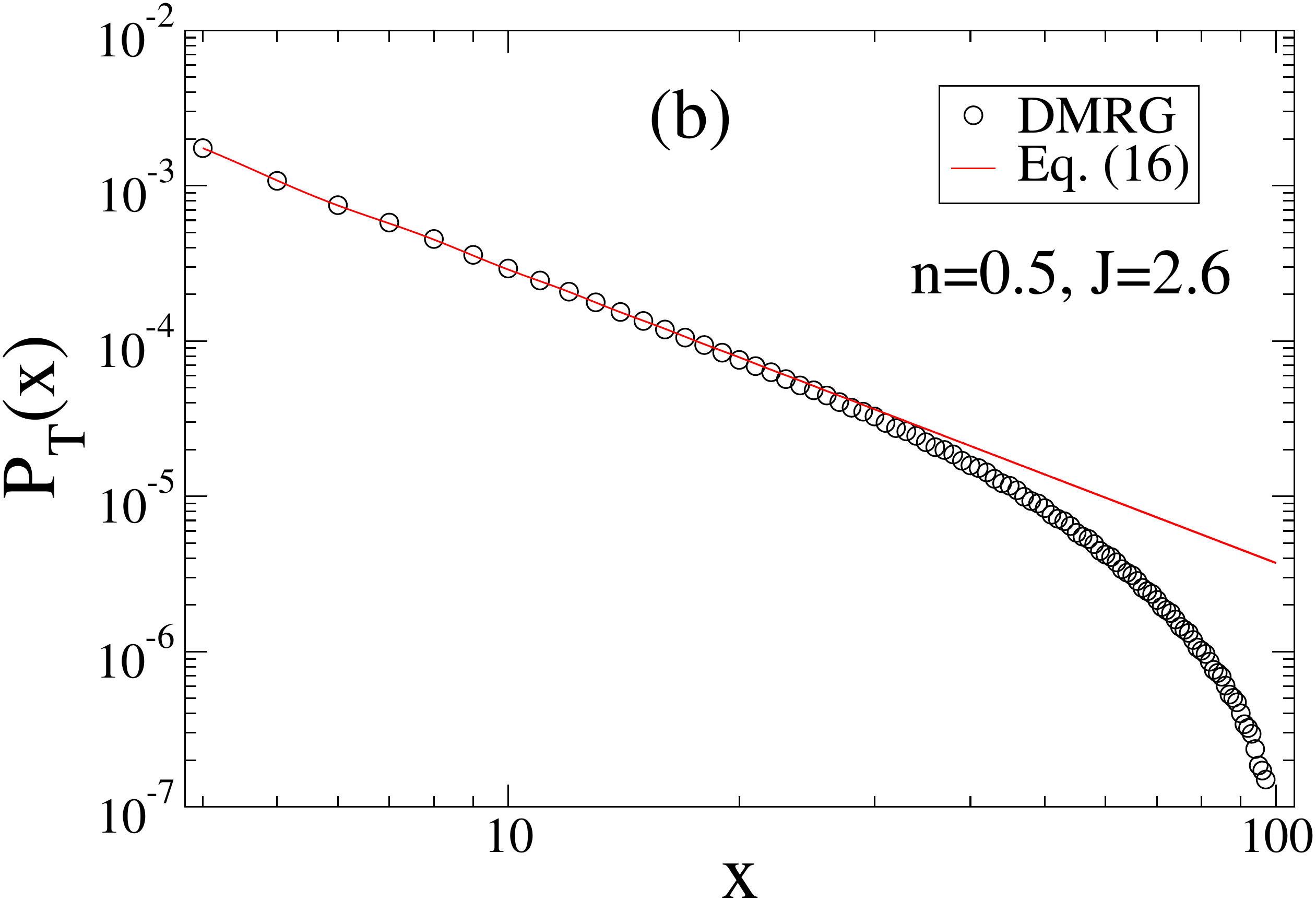}}
\end{array}$
\end{center}
\caption{(color online). Singlet $P_S (x)$ and triplet $P_T (x)$ correlation functions for $n=0.5$, $J=2.6$ and $L=200$. For the singlet case we plot $|P_S (x)|$. The straight lines correspond to power-laws determined by $K_\rho$. 
Both the singlet and triplet channel have the same exponent.} 
\label{PxST}
\end{figure}
Figure \ref{PxST} shows the behavior of both correlation functions in real space as compared to the asymptotic forms given in Eqs.\ (\ref{PScorr}) and (\ref{PTcorr}). In Fig.\ \ref{PxST} (a) a comparison is given with Eq.\ (\ref{PScorr}), where the constants $C_0$ and $C_1$ were adjusted through a least-square fit, while the value for $K_\rho$ was taken from the determination detailed in Sec.\ \ref{MetallicPhase} ($K_\rho =1.12$ for the parameter values in Fig.\ \ref{PxST}). In order to be able to display the power-law behavior of $P_S (x)$, we use a doubly logarithmic scale, and actually plot the modulus of $P_S(x)$, since the $2 k_F$ oscillations lead to sign changes of the correlation function. The dashed line through the maxima makes furthermore evident, that the power-law decay can be well described by the power of the first term in Eq.\ (\ref{PScorr}). Deviations are observed for $x > 60$, possibly due to boundary effects.
On the other hand, Fig.\ \ref{PxST} shows that the decay of $P_T$ can be well described by the same power-law as for $P_S (x)$, with deviations that start at $x > 30$. 
\begin{figure}[th!]\relax
\centerline{\includegraphics[width=3in]{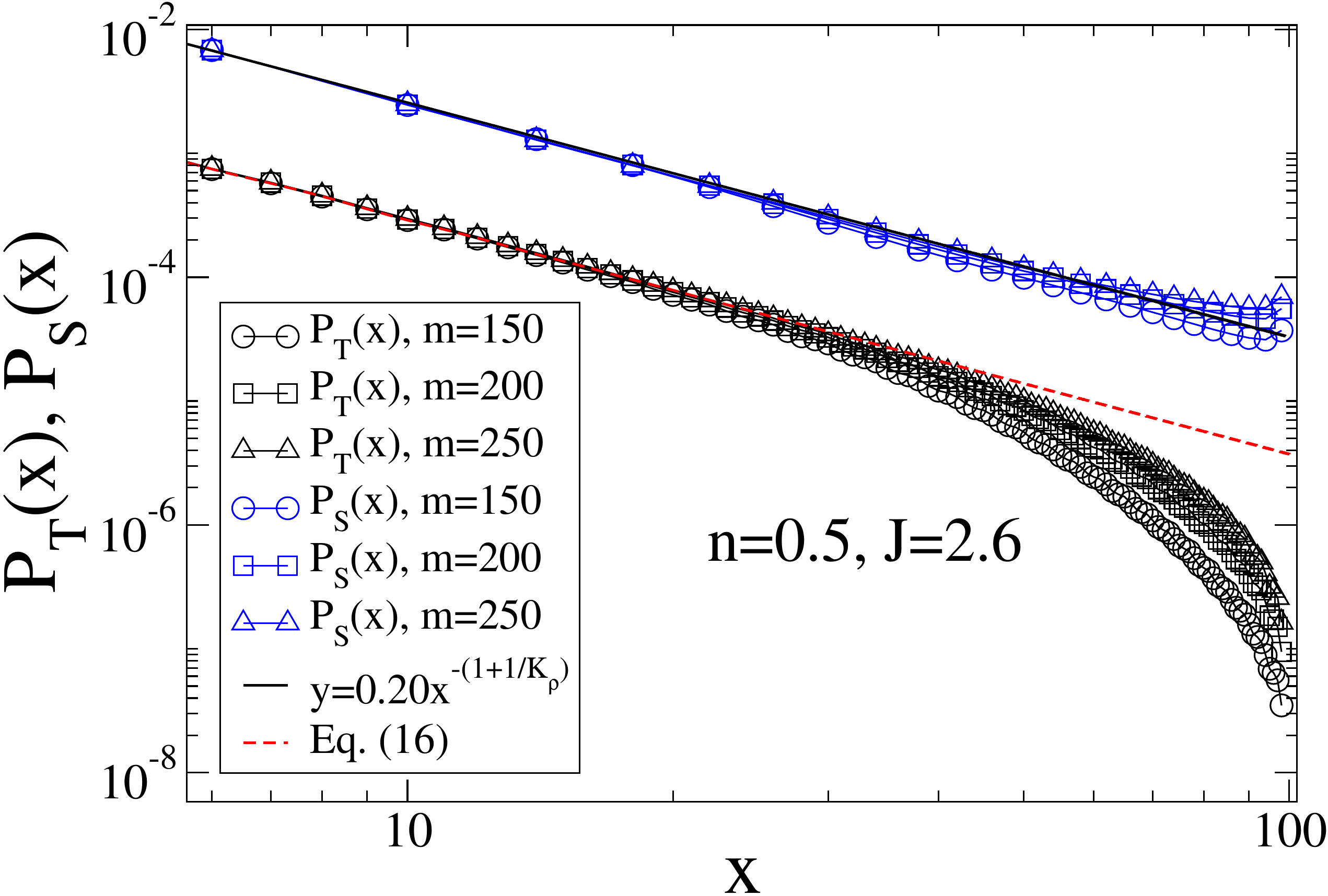}}
\caption{(color online). Singlet $P_S (x)$ and triplet $P_T (x)$ correlation functions for $n=0.5$, $J=2.6$, $L=200$ and $m=150,200,250$. We observe at long distances that 
on a double-logarithmic scale, $P_T (x)$ becomes more linear upon increasing $m$, i.e., the spurious exponential decay introduced by the DMRG 
cutoff
decreases. 
Note, however, that $P_S (x)$ becomes less linear at long distances when increasing $m$. We associate this behavior to the open boundary conditions used.
} 
\label{PxTS_ms}
\end{figure}
While there are certainly boundary effects, Fig.\ \ref{PxTS_ms} shows that their incidence on the correlation functions changes by increasing the precision of the DMRG runs. There we display the results for both correlation functions when the number $m$ of states kept in the reduced density matrix is increased. While $P_T (x)$ approaches the power-law at larger distances, $P_S (x)$ departs from it. At the highest accuracy used, both depart from the predicted power-laws for $x > 50$. 

\begin{figure}[ht!]
\centerline{\includegraphics[width=3in]{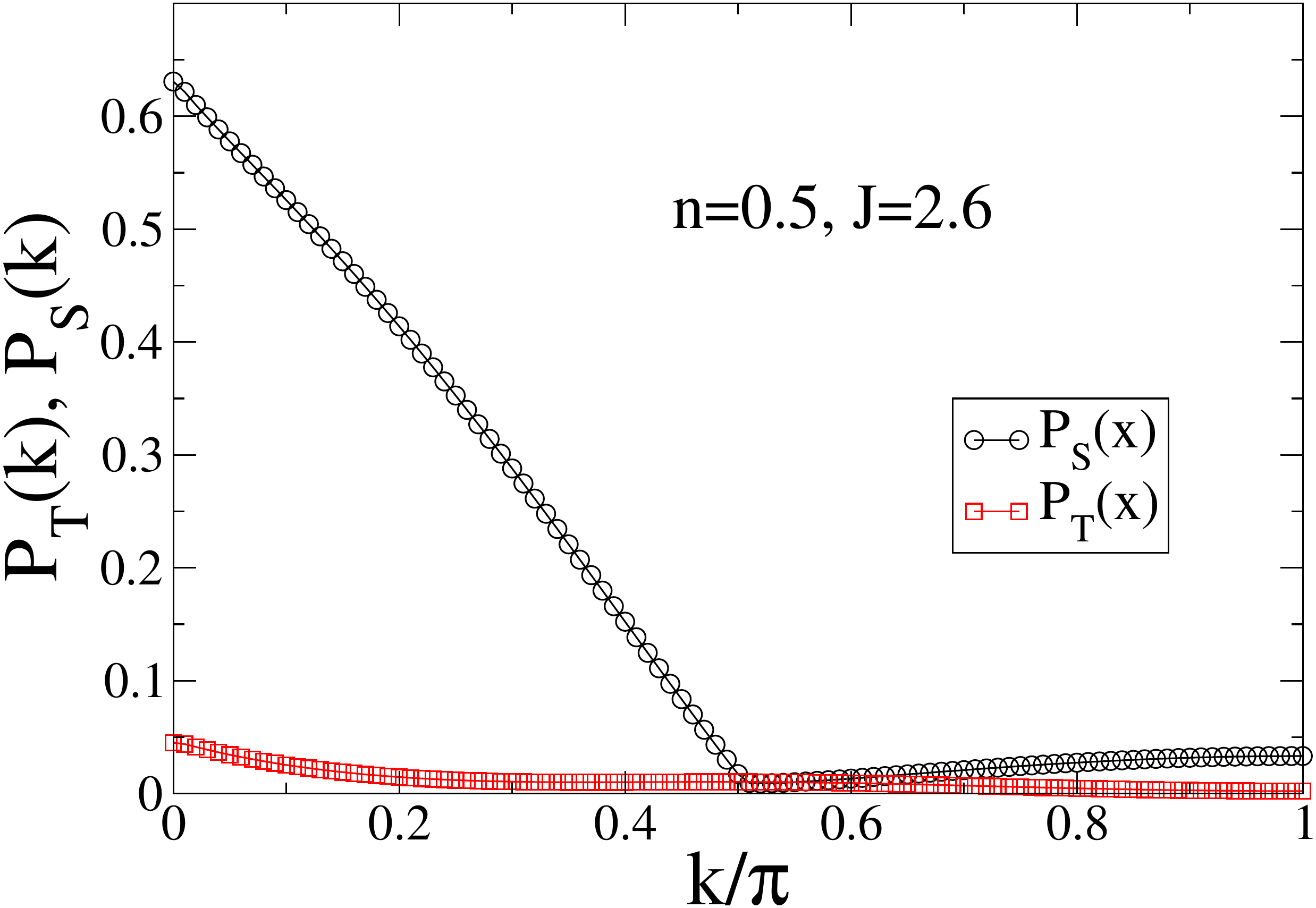}}
\caption{(color online). Triplet and singlet pair-pair structure factors for $n=0.5$, $J=2.6$ and $L=200$.} 
\label{PkST}
\end{figure}
In Fig.\ \ref{PkST} we also show both singlet and triplet pair correlations in momentum space. The amplitude of $P_S (k=0)$ for the singlet case is one order of magnitude larger than $P_T (k=0)$ for the triplet case. 
Since $P_S(k=0)$ gives the number of pairs with momentum zero, it is clear that  
singlet superconductivity dominates.

\subsection{Competing orders in the spin-gap phase. Charge density wave 
vs singlet superconductivity 
}
\label{CDW_vs_SS}
The peaks of Fig.\ \ref{Nk} (a) at $2 k_F$ and Fig.\ \ref{Pk} (a) at $k=0$ for $J = 2.5$ show that CDW and SS are competing orders in the spin-gap phase. On entering this phase, electrons pair into singlets, such that
the spin-spin and triplet pair-pair correlation functions are exponentially suppressed. While the singlet pair-pair correlation function has the long-distance behavior given by Eq.\ (\ref{PScorr-gap}), the density-density correlation function in the LE phase behaves as 
\cite{giamarchi04}
\begin{eqnarray}
\langle n(r)n(0) \rangle  & = & A_0 r^{-2}
+ A_1 \cos(2k_F r)r^{-K_\rho}.
\label{ncorr-gap}
\end{eqnarray}
Since in the spin-gap region $K_\rho > 1$,  
SS is the dominant order. 
\begin{figure}[th!]\relax
\centerline{\includegraphics[width=3in]{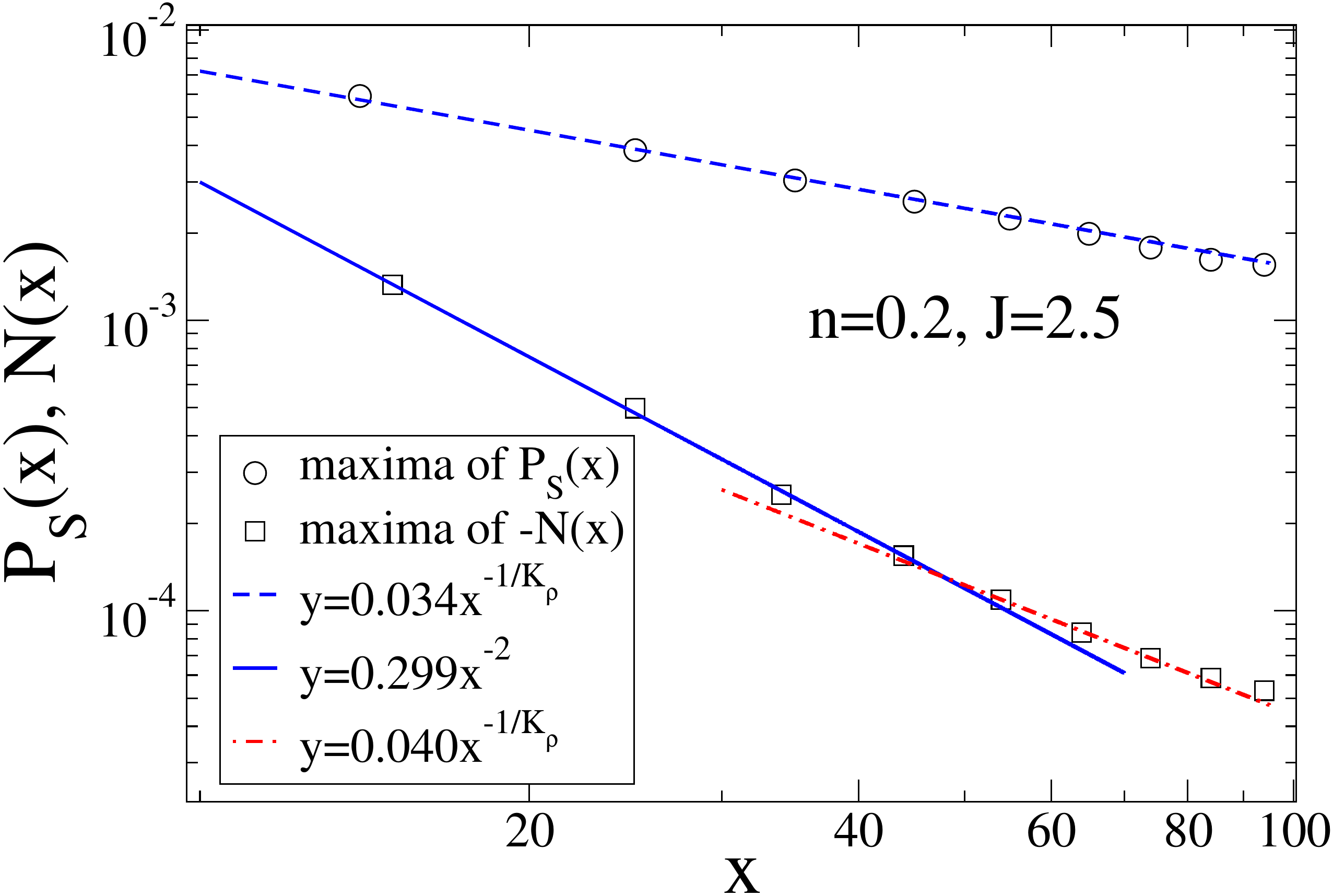}}
\caption{(color online). Maxima of $P_S (x)$ and $-N(x)$  in real space for $L=200$, $n=0.2$ and $J=2.5$ on a log-log scale. the power-laws are determined by $K_\rho$ through Eqs.\ (\ref{PScorr-gap}) and
(\ref{ncorr-gap}).}
\label{Px_Nx}
\end{figure}
The difference between both correlation functions is displayed in Fig.\ \ref{Px_Nx}, where 
we consider them at $n=0.2$ and $J=2.5$, i.e.\ deep in the spin-gap region. Due to the $2 k_F$ oscillations, only the maxima of the functions are plotted.
The different lines (dashed one for $P_S (x)$ and full and dashed-dot lines for $N (x)$) correspond to the powers appearing in the first term of Eq.\ (\ref{PScorr-gap}), and the powers appearing in Eq.\ (\ref{ncorr-gap}) for $N(x)$. The
latter shows a crossover from a behavior at short distances dominated by the first term in 
Eq.\ (\ref{ncorr-gap}) to the long-distance behavior determined by the power of the second term. For the parameters considered here we obtained from $N(k \rightarrow 0)$, as given by Eq. (\ref{e6}), $K_\rho = 1.48$.
\begin{figure}[th!]\relax
\centerline{\includegraphics[width=3in]{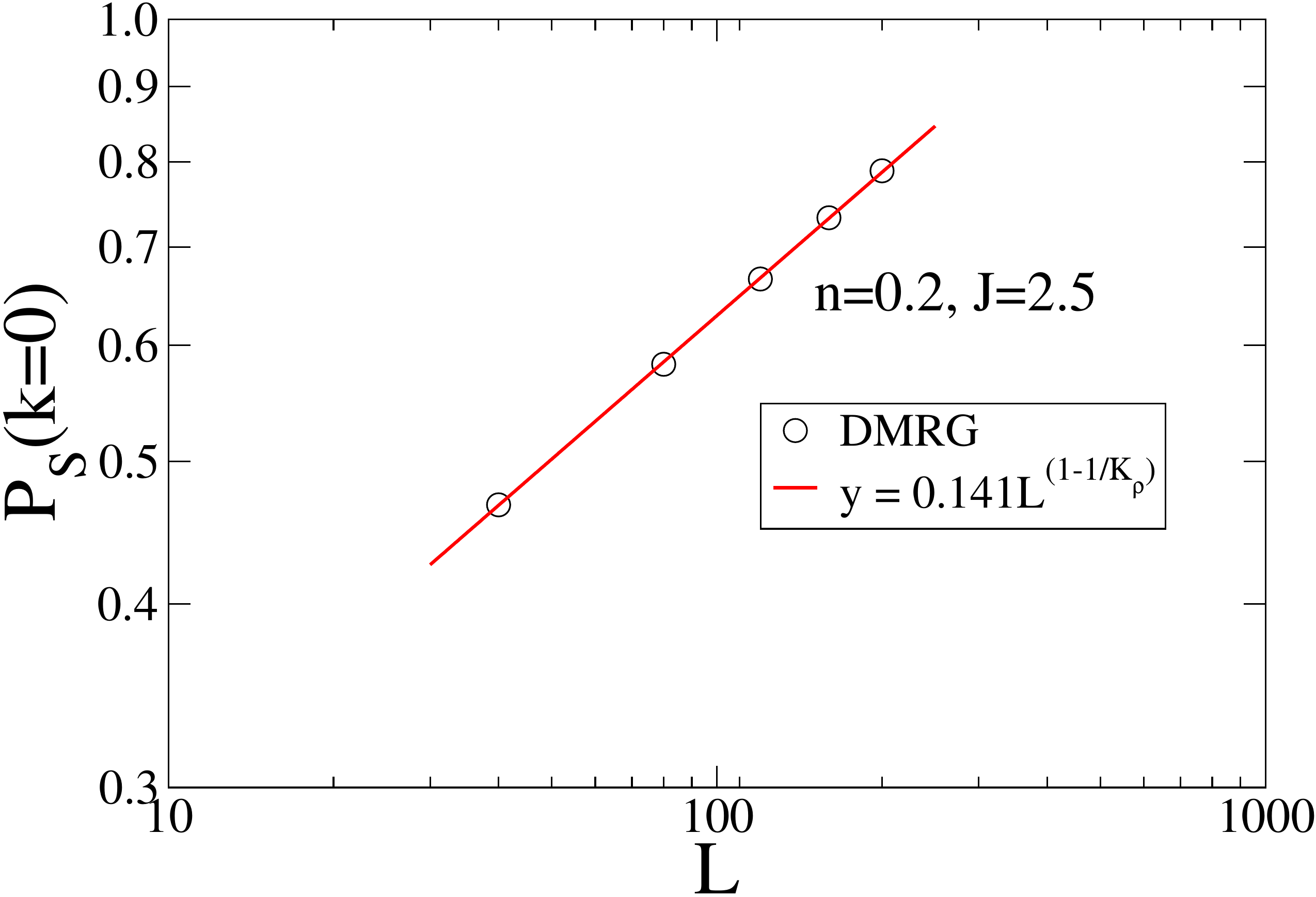}}
\caption{(color online). Scaling of $P_S(k=0)$ in the thermodynamic limit for $n=0.2$ and $J=2.5$ (spin-gap phase).}
\label{Pk_scaling}
\end{figure}
In this case $P_S(k=0)$ diverges in the thermodynamic limit, as shown by Fig.\ \ref{Pk_scaling} with an exponent determined by $K_\rho$, however, the system being one-dimensional, no true long-range order
is present, since $P_S(k=0)/L$ vanishes in the thermodynamic limit.  
Nevertheless, as shown in Fig.\ \ref{gap_vs_rho}, the spin gap increases as the density decreases, reaching a sizable 
value on approaching phase separation ($\Delta E_S \sim 0.15 t$). Hence, singlets bind at a finite temperature scale below the spin gap. 

\begin{figure}[th!]\relax
\begin{center}
$\begin{array}{c}
\includegraphics[width=3in]{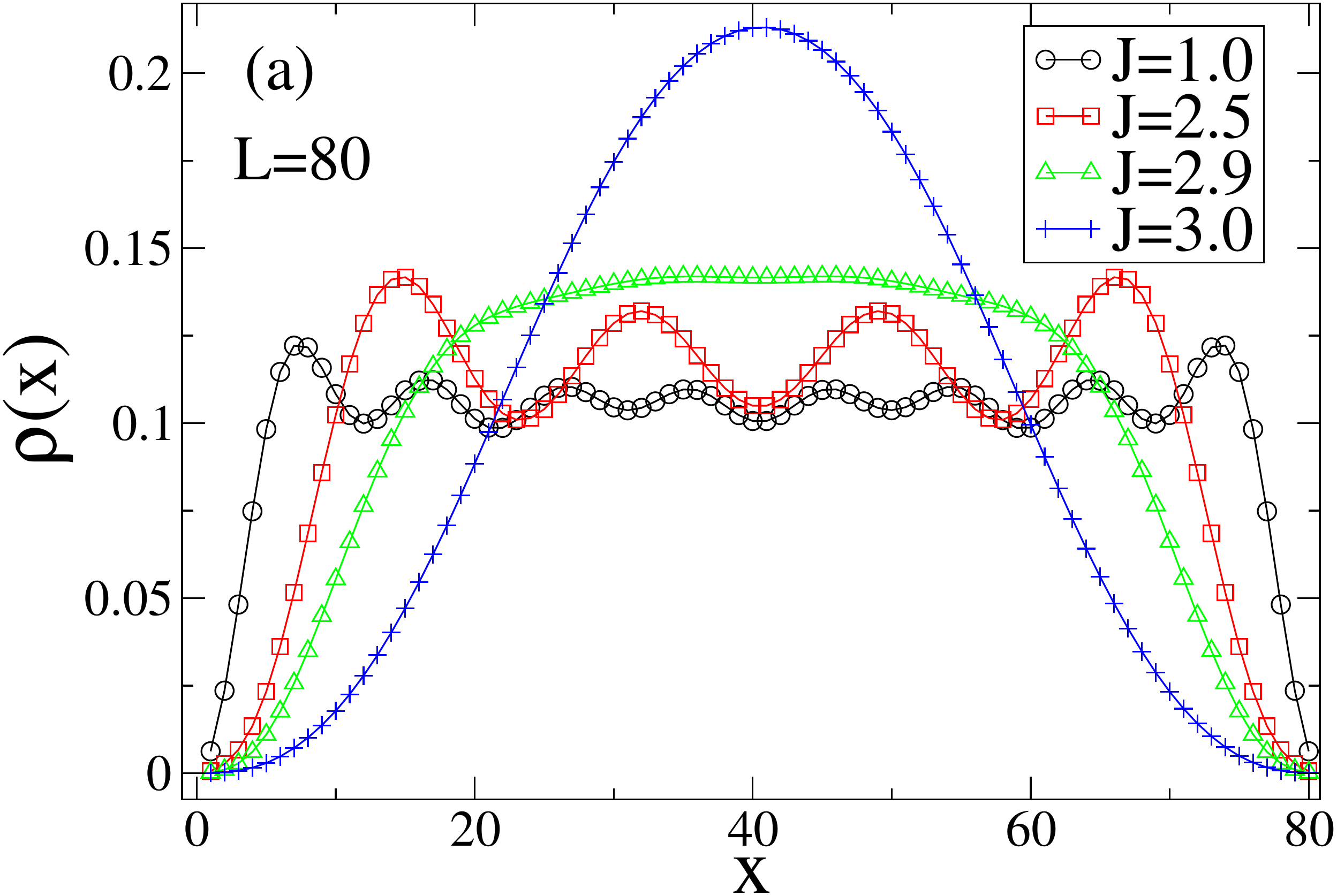}  \\ 
\includegraphics[width=3in]{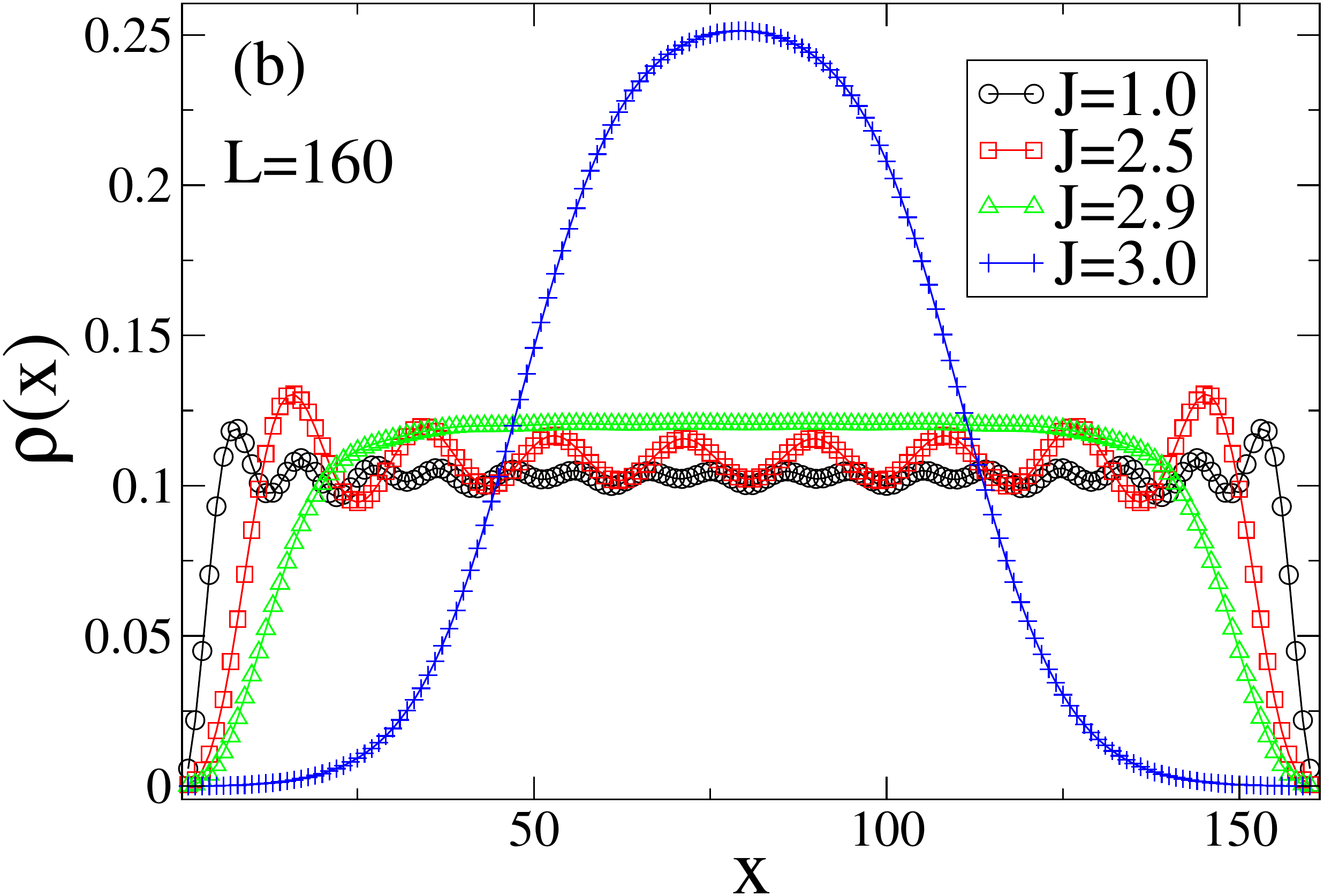} \\
\end{array}$
\end{center} 
\caption{(color online). Density in real space $\rho (x)$ for $n=0.1$ and (a): $L=80$, (b): $L=160$. In the metallic regime ($J=1.0$) we have one peak per particle due to Friedel oscillations. When $J$ is increased the particles start to form pairs. This happens in the spin-gap region ($J=2.5$). Increasing $J$ even more the particles tend to be confined in a region smaller than the system size (Phase separation).}
\label{dens_x}
\end{figure}

Another interesting aspect of this phase is that due to open boundary conditions, a change in the periodicity of the Friedel oscillations in the density can be directly observed, as shown in Fig.\ \ref{dens_x}.
This can be seen as a $2 k_F$ CDW coexisting with a superconducting state in the spin-gap phase. 
However, following the arguments related to Fig.\ \ref{gas_bound_ene}, and the fact that the leading singularity is related to SS, these density oscillations can be understood
as due to bound pairs, which because of the constraint of the t-J model behave as hard-core bosons. Hence, the density of the hard-core bosons in one dimension will show the same oscillations as the density of the equivalent Jordan-Wigner fermions, corresponding to the number of pairs in the system. As an example we show in Fig.\ \ref{dens_x} the density profiles for a density $n=0.1$ and values of $J$ corresponding to the repulsive LL ($J = 1$) and the spin-gap ($J = 2.5$) phases, and two values of $J$ on entering the phase separation, for two different system sizes (Fig.\ \ref{dens_x} (a): $L=80$ and Fig.\ \ref{dens_x} (b): $L=160$). In both cases, the number of oscillations is halved, corresponding to Friedel oscillations for half the number of particles.

\begin{figure}[ht!]
\begin{center}
$\begin{array}{c}
\includegraphics[width=2.8in]{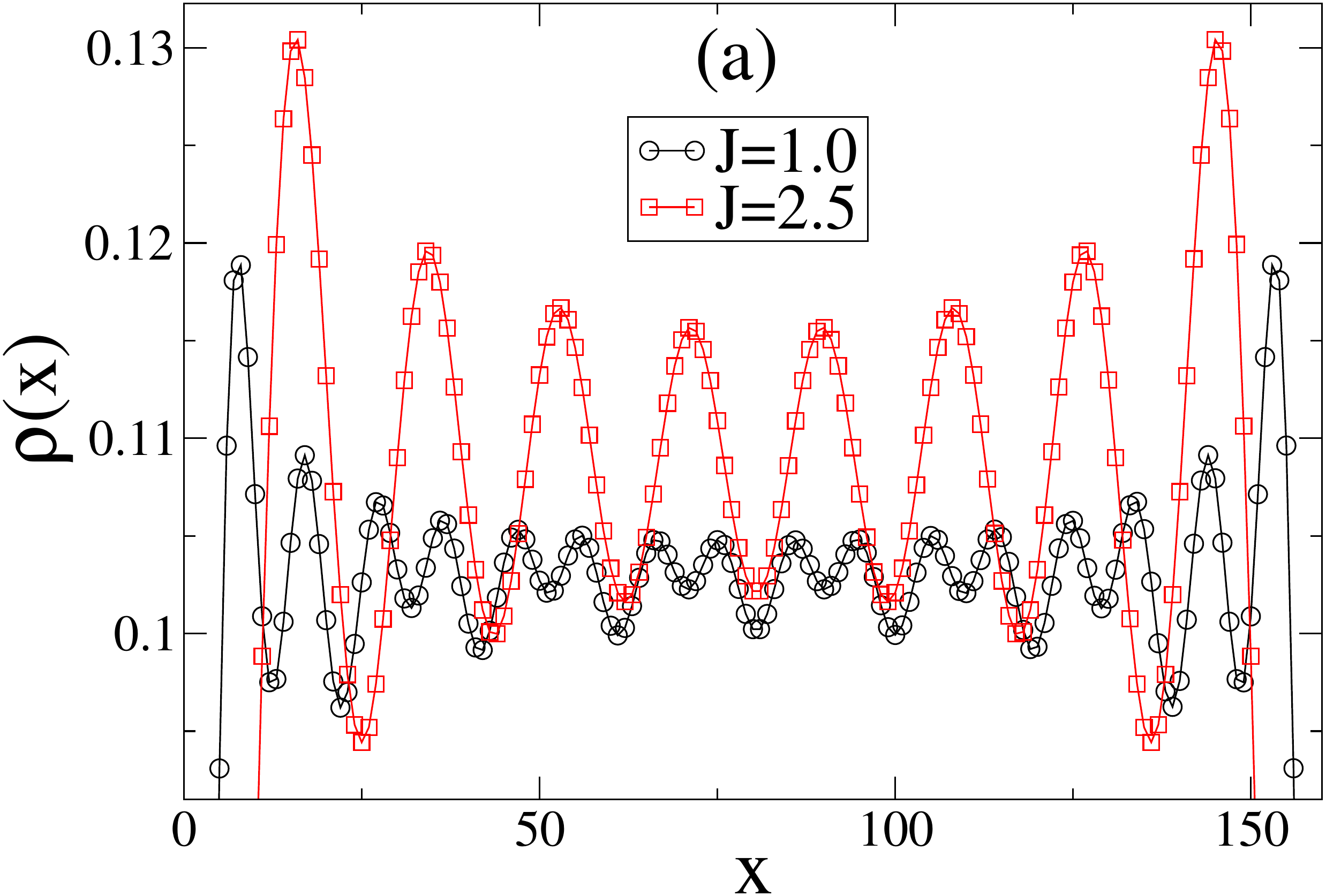}  \\ 
\includegraphics[width=2.8in]{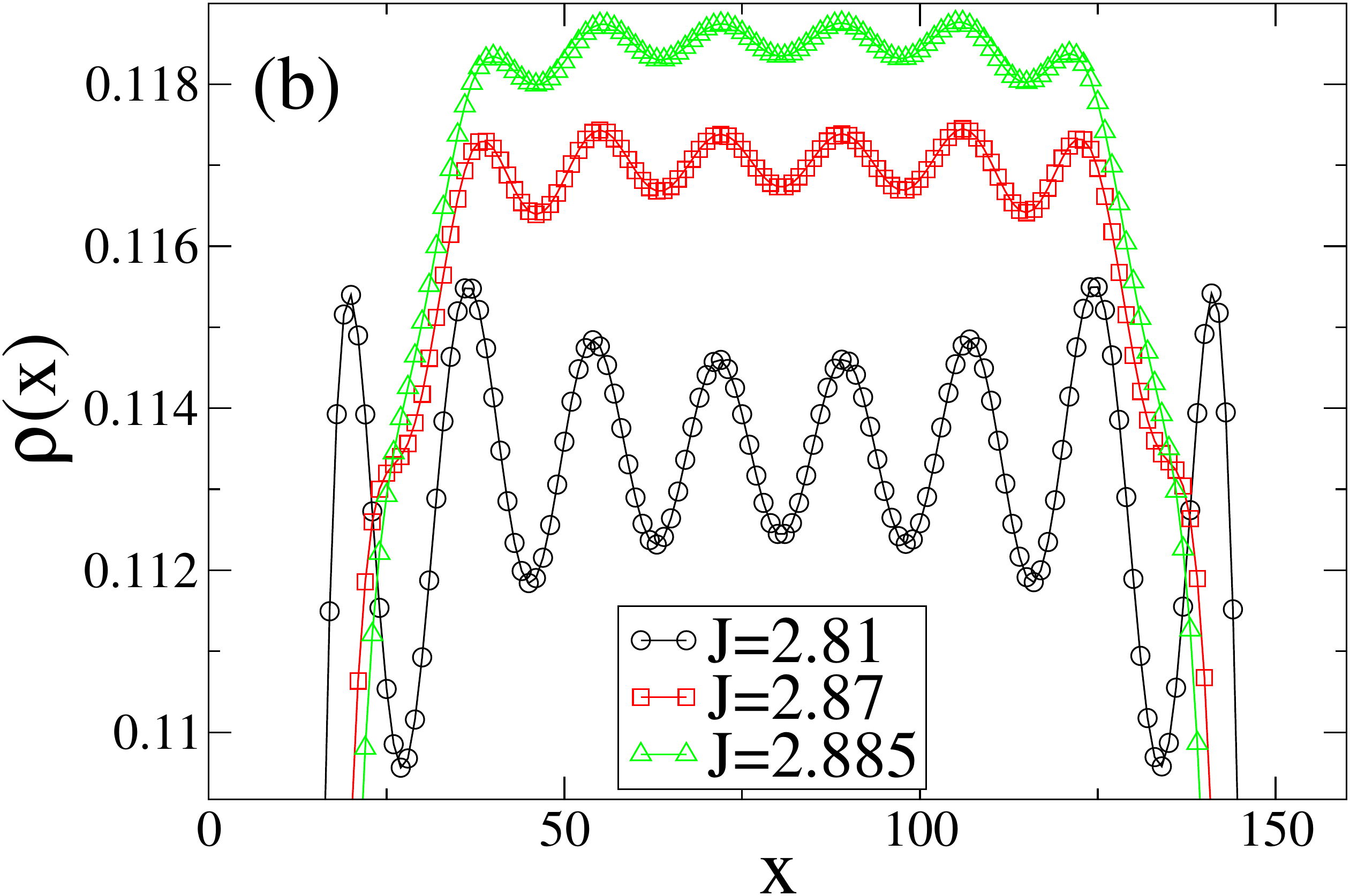} \\
\includegraphics[width=2.8in]{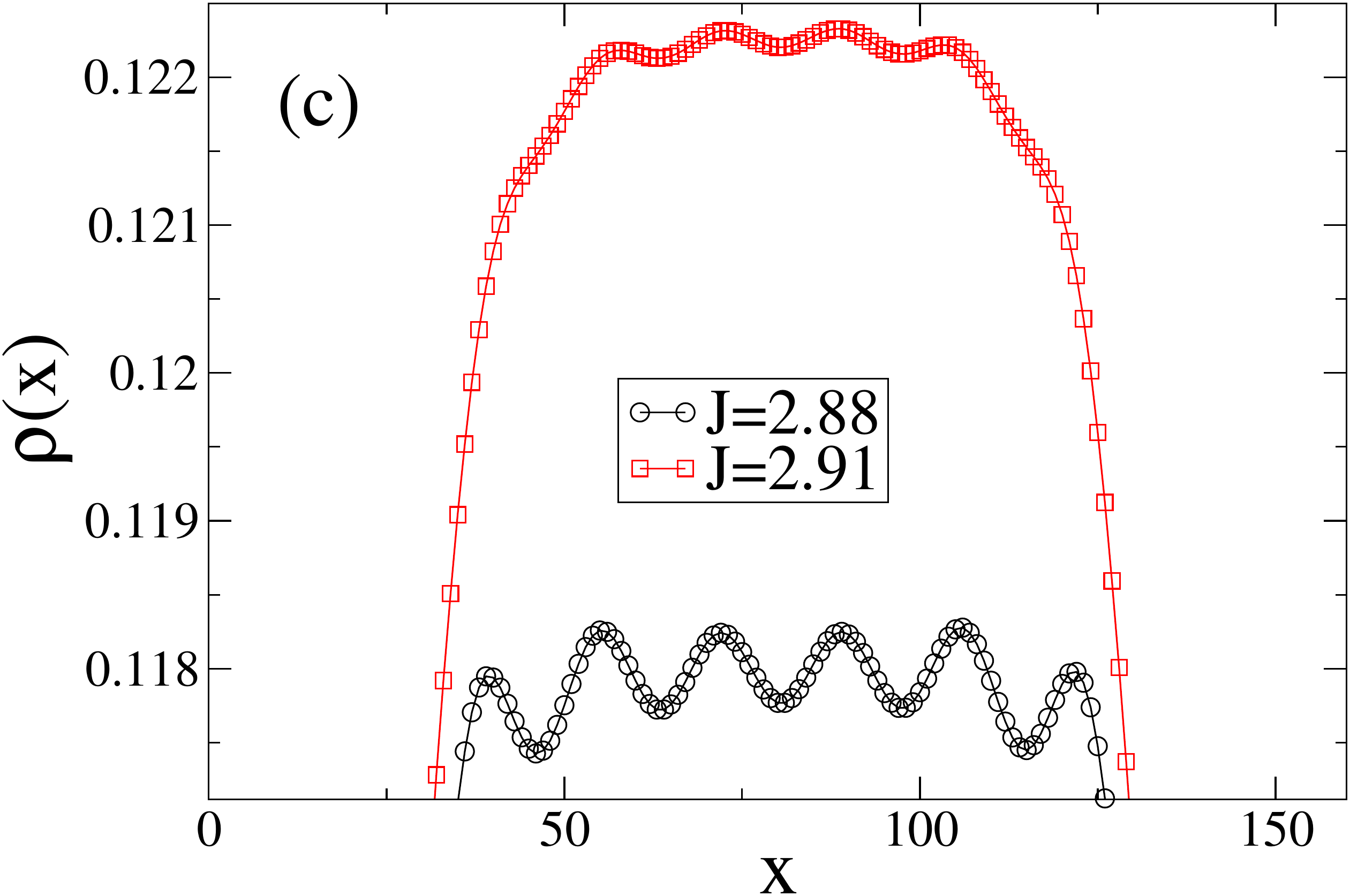}  \\ 
\includegraphics[width=2.8in]{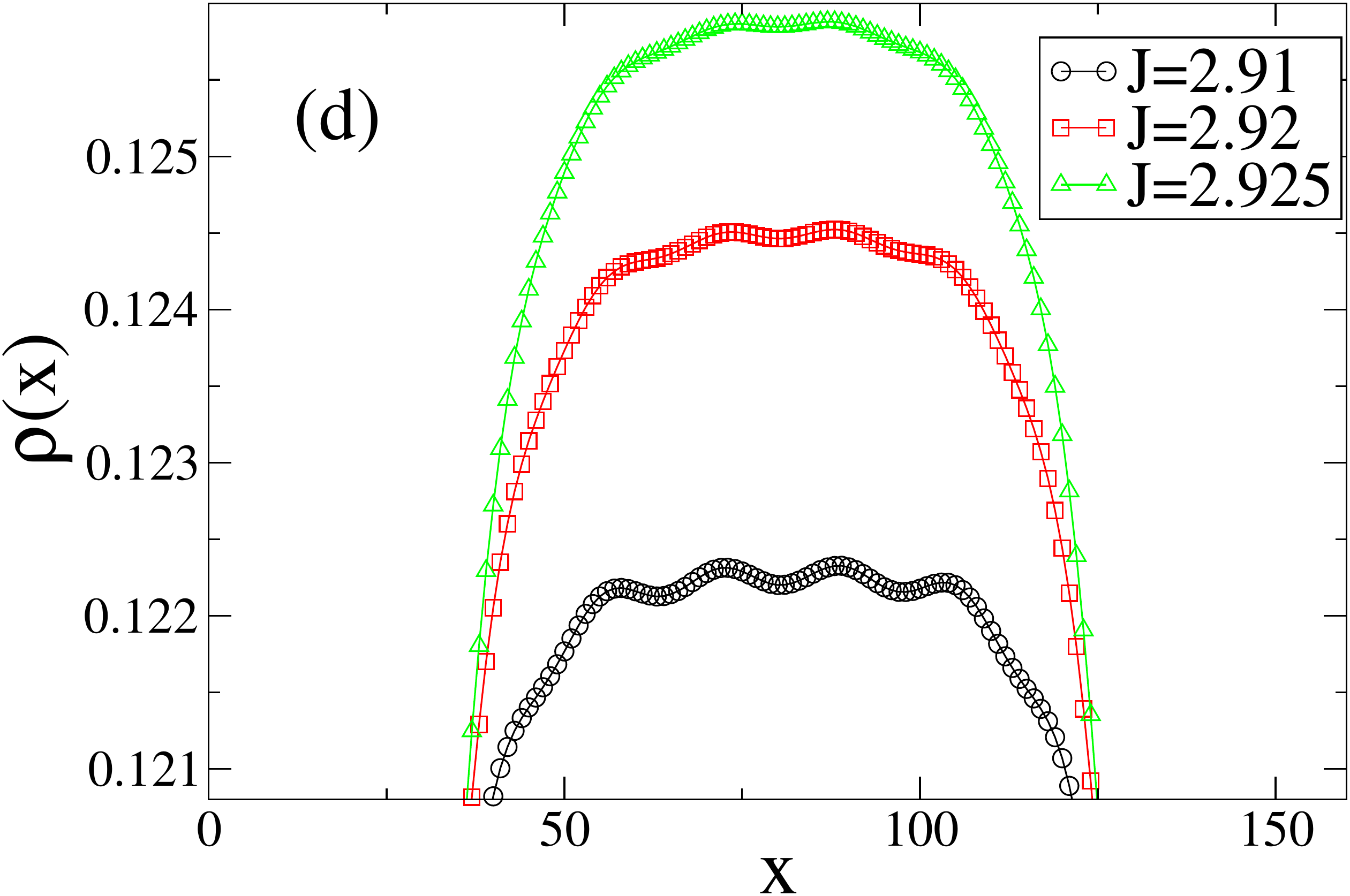}  \\
\end{array}$
\end{center}
\caption{(color online). Density in real space $\rho (x)$ for $n=0.1$, $L=160$ and different values of $J$. In Fig.\ \ref{condensation}(a) we observe the pairing of particles. In the other figures only particles at the boundaries of the particle-rich region merge into one loose cloud but the particles in the middle do not cluster further.}
\label{condensation}
\end{figure}
Since the coupling driving the pairing mechanism
is the nearest neighbor exchange $J$, it could be argued \cite{ogata91} that before entering phase separation, clustering in groups with more than two fermions may exist.
In order to answer this question we observe how the modulations of $\rho (x)$, displayed in Fig.\ \ref{condensation}, change when we increase $J$ by very small amounts before phase-separation is reached.
There, we clearly see how a loosely bound cloud forms at the border of the particle rich region, but pairs remain as such in the middle of that region. Starting with 16 electrons on 160 sites, we can see the formation of pairs on entering the spin-gap region, such that 8 pairs are clearly visible (Fig.\ \ref{condensation} (a)). Going up to the point where density oscillations corresponding to two pairs are still visible (Fig.\ \ref{condensation} (d)), we see that the rest of the pairs merged on the sides of the particle rich region, until phase-separation is reached. Note that the density oscillations in Figs.\ \ref{condensation} are on a very small scale and are not noticeable at the scale used  in Figs.\ \ref{dens_x}. Once in the 
phase-separated region, a
density distribution results, where a cloud at the center of the system appears, leaving an appreciable number of empty sites, as shown in Figs.\ \ref{dens_x}.
Hence, clusters with more than a pair do not form a uniform phase, and in particular, a state with four bound electrons does not form.
Still inside the phase-separated region, appreciable changes of the density as a function of $J$ take place. This will be the subject of the next section.

\subsection{Phase separation and electron solid phase \label{PairingRS}}
\begin{figure}[th!]\relax
\begin{center}
$\begin{array}{c}
\includegraphics[width=3in]{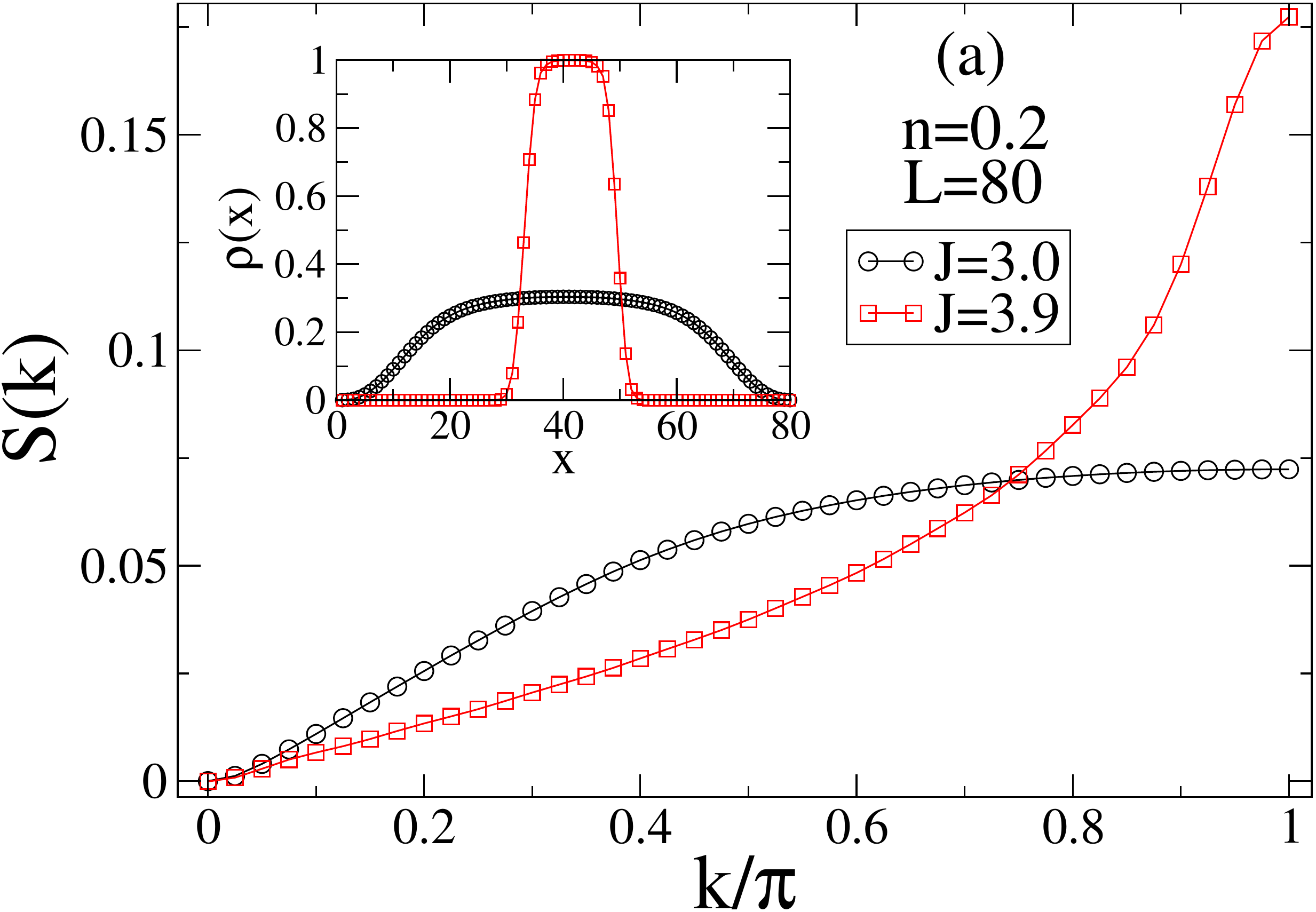}  \\ 
\includegraphics[width=3in]{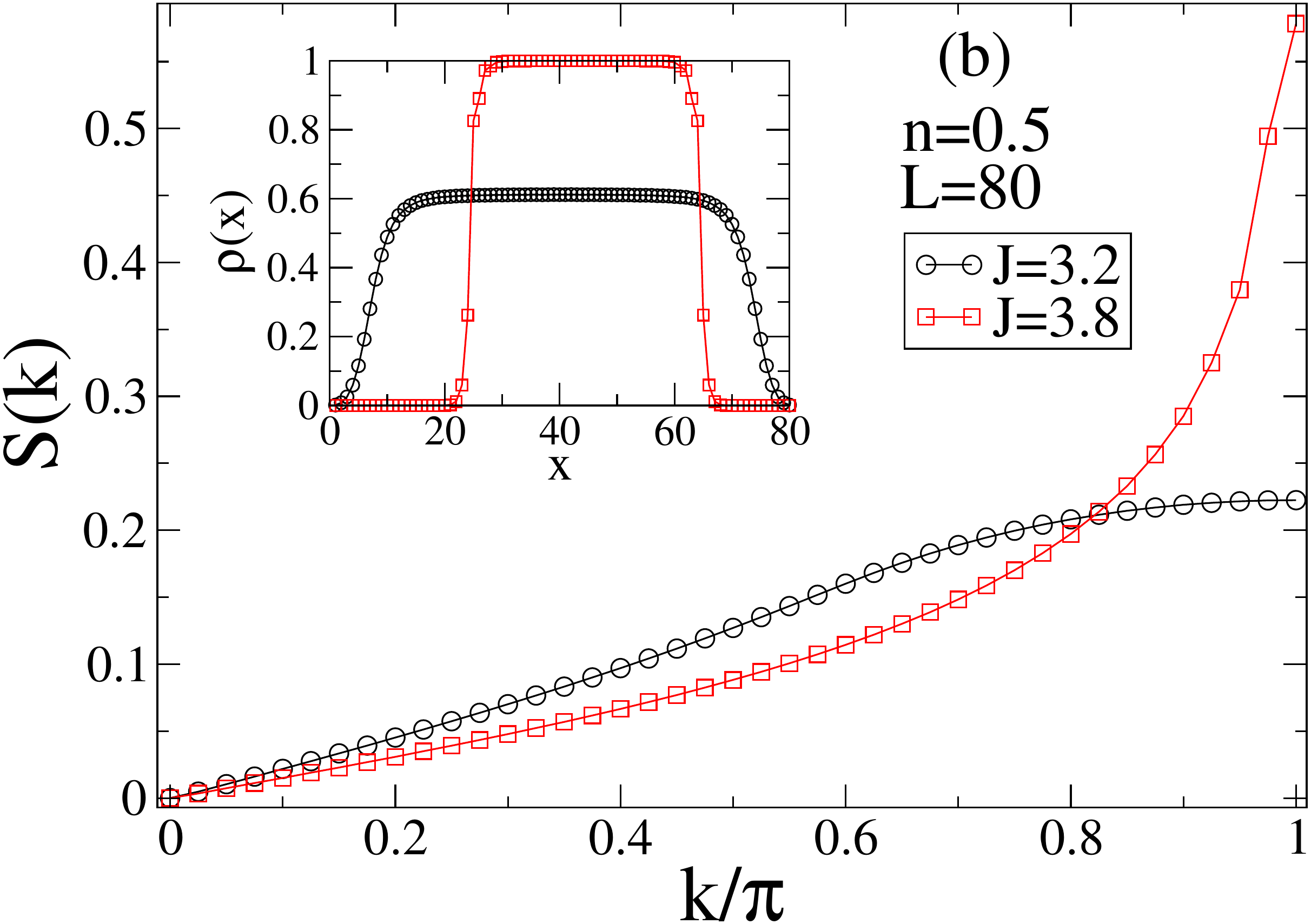} \\
\includegraphics[width=3in]{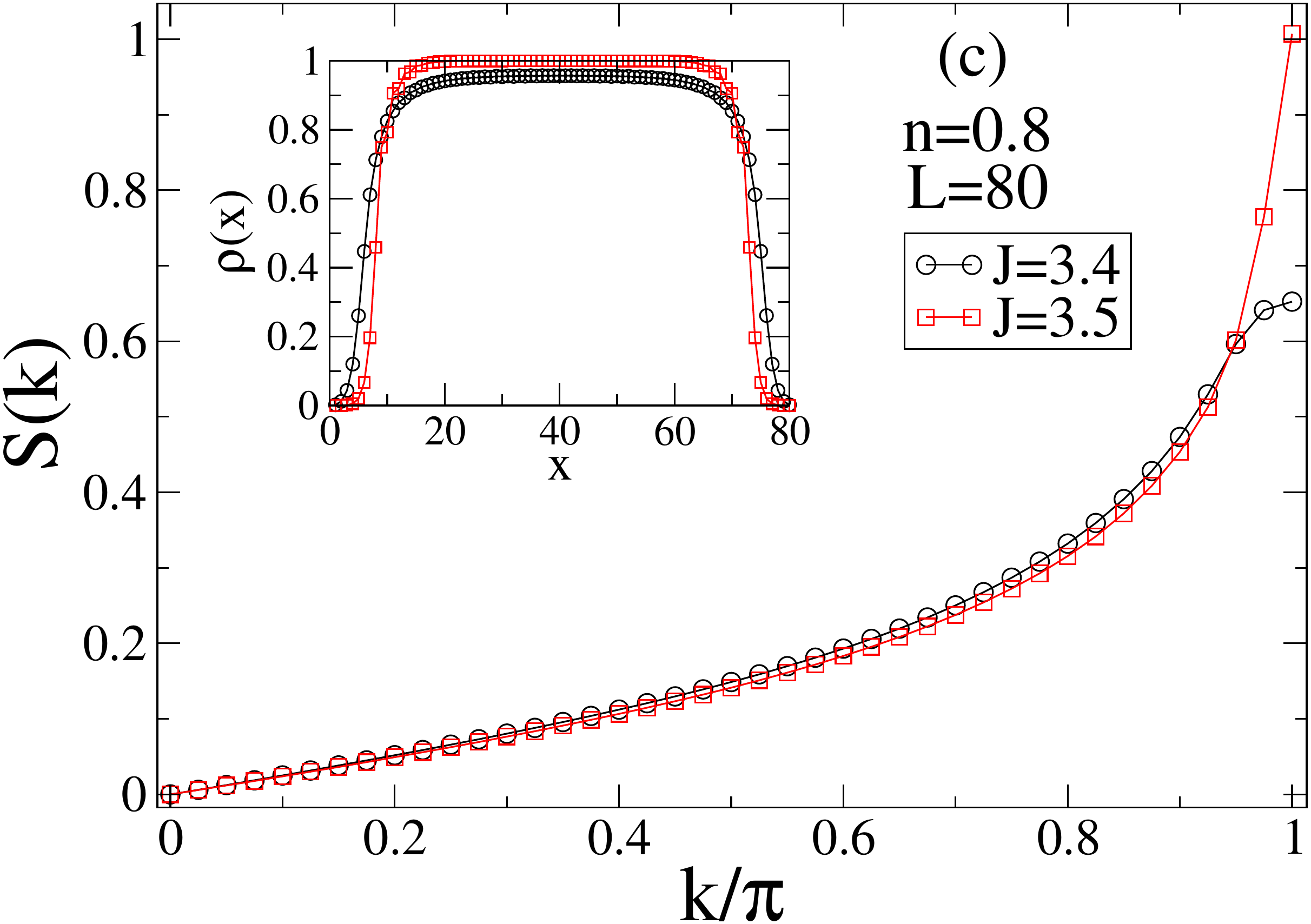}
\end{array}$
\end{center} 
\caption{(color online) Magnetic structure factor $S(k)$ in the phase-separated region for values of $J$ close to the boundary and at the value, where a peak at $k=\pi$ emerges. Insets: the corresponding density profiles}
\label{SkinPhaseSeparation}
\end{figure}
As shown in Fig.\ \ref{dens_x} at values of $J$ 
just after the phase boundary ($J=3$), particles merge into a single island, occupying only part of the available space. However, as shown by the magnetic structure factor in Fig.\ \ref{Sk}, for values of $J$ inside phase-separation, and close to the boundary, a broad maximum is seen at $k=\pi$, but no sharp peak indicative of the formation of a spin chain. Therefore, the na\"ive picture of a compact region does not apply yet. 
Figure \ref{SkinPhaseSeparation} shows $S(k)$ at three different densities for the values of $J$ where the system enters phase-separation and the one at which a sharp maximum can be observed at $k = \pi$. The insets display also the corresponding density profiles, making evident that when $S(k)$ has a sharp maximum at the antiferromagnetic wavevector, an island with density $n=1$ is formed. Unfortunately, since the formation of such islands implies that there are many almost degenerate states very close to the ground-state, namely those connected by translation, is is not possible to perform a careful finite-size analysis to determine the boundary to such a phase in the thermodynamic limit. In fact, in Figs.\ \ref{SkinPhaseSeparation} it can be seen that the density profiles reaching $n=1$ break spontaneously reflexion symmetry about the central bond, an artifact due to the many degenerate states mentioned above.

\begin{figure}[hb]\relax
\begin{center}
\includegraphics[width=3in]{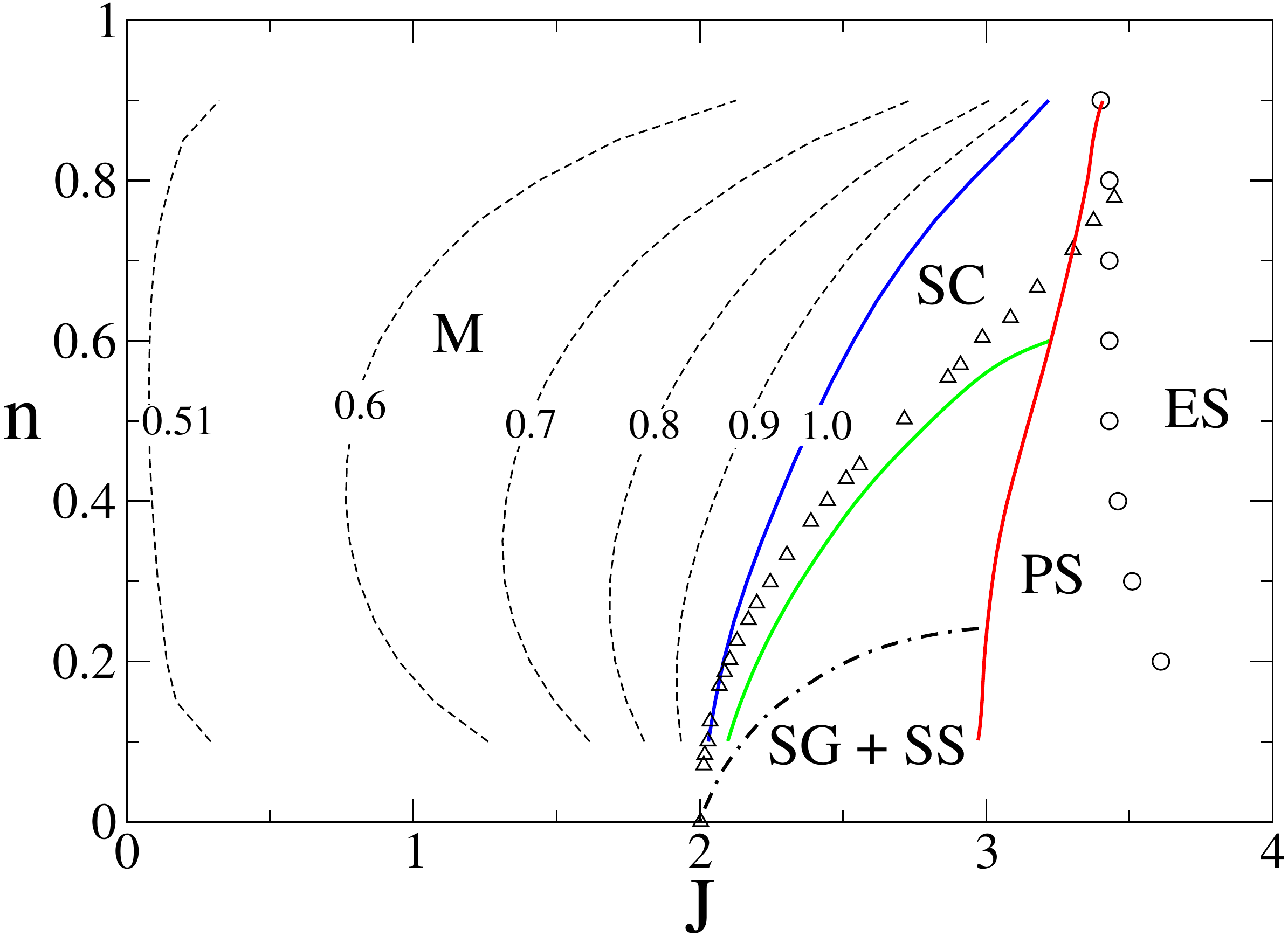} 
\end{center} 
\caption{(color online) 
Phase diagram of Fig.\ \ref{fig1} including the electron solid phase (ES).
The circles denote the position at which a peak in $S(k)$ appears at $k=\pi$, and at the same time, the phase separated island reaches a density $n=1$. 
Triangles reproduce the results of Ref.\ \onlinecite{nakamura97} and the dashed-dot line those of Ref.\ \onlinecite{helberg93}.}
\label{LastPhaseDiagram}
\end{figure}
The difference in density profiles shown in Figs.\ \ref{SkinPhaseSeparation} makes also evident that by having an expanded cloud, most probably with pairs inside, as discussed in the previous section, still kinetic energy is present in the cloud that is quenched only when $J$ reaches a large enough value. 
Since we are limited in the sizes of the systems we can simulate, we cannot determine with confidence a critical value of $J$, where the system forms an electron solid\cite{chen96}. It was argued\cite{chen96} that in this case $J_c$ should be independent of the band filling, $n$, 
since the only requirement is to form an island with $n=1$.  We observe in our simulations a weak dependence on the density, but our results are limited to $L=80$, the largest size where meaningful results can be obtained in this region. Placing the location of the value of $J$, where the island with $n=1$ forms for $L=80$ at different densities, leads to an almost density independent value, as shown in Fig.\ \ref{LastPhaseDiagram}.
The largest deviations are observed at the lowest densities. Although
the number of particles is possibly too low to make a definite statement, it could be argued that at low densities the formation of the electron solid leads to an appreciable loss of kinetic energy, that has to be compensated by a larger value of $J_c$.  The same can be argued in the high density region, such that being limited to small systems may lead to lower values of $J_c$, since, as displayed in Fig.\ \ref{SkinPhaseSeparation} (c), for very few holes the loss in kinetic energy with open boundary conditions 
can be negligible.

Figure \ref{LastPhaseDiagram} also displays the differences in the results for the onset of the spin-gap region that were obtained on the basis of a variational calculation combined with the power method \cite{helberg93} on the one hand, and a RG analysis complemented by exact diagonalizations of systems with up to 30 sites \cite{nakamura97}, on the other hand. While the agreement with the latter results is in general quite good, the largest deviations appear as the onset line approaches the phase-separation line. This can be possibly due to the fact that the RG analysis did not take explicitly into account the appearance of phase-separation, that in general leads to a strong finite size dependence. On the other hand, a precise determination with DMRG becomes extremely demanding at low densities ($n \leq 0.1$), due to the need of very large systems in order to have a large enough number of particles to faithfully represent a phase. The large disagreement with the results of Ref.\ \onlinecite{helberg93} could point to the difficulties of the power method to deal with 
fermionic systems at high density, due to the strong increase of the dimension of the required Hilbert space.

\section{Summary}

We have revisited the phase-diagram of the one-dimensional t-J model and determined, on the basis of finite-size extrapolations of results obtained with DMRG, the boundaries between the known four phases: 
metal (M), singlet-superconductivity with spin gap (SG+SS), gapless superconductivity (SC), and phase separation (PS) (see Fig.\ \ref{fig1}) . The most controversial issue was related to the boundary between SC 
and SG+SS, where appreciable differences were present between the results from 
variational methods \cite{chen93,helberg93}, and results based on renormalization group
\cite{nakamura97}. The highest densities at which the spin-gap phase was predicted was $n \sim 0.4$ and $n \sim 0.8$, respectively. In our case it corresponds to $n \sim 0.6$. The boundary between M and SC was determined by extracting the Luttinger liquid anomalous dimension $K_\rho$ from the slope of the structure factor for density correlations in the limit $k \rightarrow 0$, extrapolated to the thermodynamic limit. The extrapolations were performed using system sizes $L = 40$, 80, 120, 160, and 200.
The opening of the spin gap was directly determined by examining the gap to the lowest triplet state, and again extrapolating to the thermodynamic limit. Finally, the boundary to PS was determined by extrapolating the inverse compressibility. 

We further characterized the different phases through correlation functions for density, spin, singlet and triplet superconductivity, and their corresponding structure factors. The correlation functions could be consistently described by the determined values of $K_\rho$. We also considered the momentum distribution function, that at high densities shows a peculiarity not possible in a single band conventional metal. Apart from detecting the $3 k_F$ singularity, at high enough densities, $n(k)$ increases for $k > k_F$. This corresponds to spectral weight for energies below the Fermi energy but $k > k_F$, as observed in previous quantum Monte Carlo simulations \cite{lavalle03}.

In the gapless superconducting phase the Luttinger parameter $K_\rho$ is larger than one and both SS and TS correlation functions decay with the same critical exponent. The amplitude of the structure factor for singlet pairing clearly dominates over that for triplet pairing, as expected (Fig. \ref{PkST}). A very good description of the correlations in real space is obtained by the forms given by Luttinger liquid theory
(Fig.\ \ref{PxST}), and with power-laws consistent with $K_\rho$ determined from the structure factor for the density-density correlation function. It describes both the decay of the correlation function for the singlet as well as the triplet channel. 

In the spin-gap phase with singlet-superconductivity we observed that the ground state energy compares very well to the ground state energy of a gas of singlet bound pairs in a region $2.0 < J < 3.0$ at very low densities (Fig. \ref{gas_bound_ene}). Correspondingly,  a $2 k_F$ singularity due to pairing can be seen in the structure factor $N(k)$ (Fig. \ref{Nk} (a) and (b)). In fact, the density profile of systems with open boundary conditions shows in this phase modulations that can be viewed as pairing in real space (Fig. \ref{dens_x}). We would like to also remark that the energy scale of the spin gap can reach values $\sim 0.1 t$, and hence their formation should be experimentally accessible  
at finite temperatures.

The expectation of clustering of electrons beyond pairs close to the boundary to phase separation as  a possible phase is not supported by our calculations. 
While increasing $J$ at low density $n$, we monitored the particle density modulations in real space (Fig. \ref{condensation}). We see that between the LL and the spin-gap phase a pairing of particles in real space in fact occurs. However, on increasing $J$ the particles build a loose cloud only at the boundaries between the particle-rich region and the hole region. In the center of the particle-rich region still modulation corresponding to pairs can be seen.

On entering phase-separation,
the following features are present: infinite compressibility (Fig \ref{fig10}), Luttinger parameter $K_\rho \rightarrow \infty$ (Fig. \ref{fig4}), and, confinement of particles in real space to a region smaller than the available space (Fig. \ref{dens_x}). 
In comparison to other studies \cite{ogata91,chen93,helberg93}, the phase-separation boundary is pushed up to higher values of $J$. As observed in early quantum Monte Carlo simulations\cite{assaad91} 
we see that the onset of phase separation and the formation of a single antiferromagnetic Heisenberg island do not 
occur simultaneously. Although a detailed finite-size scaling was precluded by metastability problems in the numerical implementation, we determined an approximate boundary to the appearance of such islands, termed previously electron solids \cite{chen96}. Figure \ref{LastPhaseDiagram}, containing these data displays the region, where such islands occur.

\acknowledgements

We are grateful to the J\"{u}lich Supercomputing Center (JSC) and the High Performance Computing Center Stuttgart (HLRS) for allocation of computer time. We acknowledge financial support by the DFG in the frame of the SFB/TRR 21, and S.\ R.\ M.\ acknowledges financial support by PIF-NSF (grant No. 0904017). A.\ M.\ and S.\ R.\ M.\ acknowledge interesting discussions with A.\ M.\ Rey and A.\ V.\ Gorshkov and A.\ M.\ is grateful to KITP, Santa Barbara for hospitality during the completion of this work. This research was supported in part by the National Science Foundation under Grant No.PHY05-51164.


\begin{thebibliography}{10}

\bibitem{zhang98}
F.~C. Zhang and T.~M. Rice, Phys. Rev. B {\bf 37},  3759  (1998).

\bibitem{chao77}
K.~A. Chao, J. Spalek, and A.~M. Oles, J. Phys. C {\bf 10},  L271  (1977).

\bibitem{ogata91}
M. Ogata, M. Luchini, S. Sorella, and F. Assaad, Phys. Rev. Lett {\bf 66},
  2388  (1991).

\bibitem{dagotto94}
E. Dagotto, Rev. Mod. Phys. {\bf 66},  763  (1994).

\bibitem{white98}
S.~R. White and D.~J. Scalapino, Phys. Rev. Lett {\bf 80},  1272  (1998).

\bibitem{eckardt10}
A. Eckardt and M. Lewenstein, Phys. Rev. A {\bf 82},  011606(R)  (2010).

\bibitem{reypriv}
A.~M. Rey and A.~V. Gorshkov, private communication  (2010).

\bibitem{schulz90}
H.~J. Schulz, Phys. Rev. Lett {\bf 64},  2831  (1990).

\bibitem{kawakami90}
N. Kawakami and S.-K. Yang, Phys. Lett. A {\bf 148},  359  (1990).

\bibitem{frahm90}
H. Frahm and V.~E. Korepin, Phys. Rev. B {\bf 42},  10553  (1990).

\bibitem{bares90}
P.~A. Bares and G. Blatter, Phys. Rev. Lett {\bf 64},  2567  (1990).

\bibitem{bares91}
P.~A. Bares, G. Blatter, and M. Ogata, Phys. Rev. B {\bf 44},  130  (1991).

\bibitem{giamarchi04}
T. Giamarchi, {\em Quantum Physics in One Dimension} (Clarendon Press, Oxford,
  2004).

\bibitem{haldane81}
F.~D.~M. Haldane, J. Phys. C {\bf 14},  2585  (1981).

\bibitem{haldane81b}
F.~D.~M. Haldane, Phys. Rev. Lett {\bf 45},  1358  (1981).

\bibitem{solyom79}
J. S\'{o}lyom, Adv. Phys. {\bf 28},  201  (1979).

\bibitem{emery79}
V.~J. Emery, {\em In Highly Conducting One-dimensional Solids} (Plenum, New
  York, 1979).

\bibitem{chen93}
Y.~C. Chen and T.~K. Lee, Phys. Rev. B {\bf 47},  11548  (1993).

\bibitem{helberg93}
C.~S. Hellberg and E.~J. Mele, Phys. Rev. B {\bf 48},  646  (1993).

\bibitem{SirkerKluemper2002}
J. Sirker and A. Kl\"umper, Phys. Rev. B {\bf 66},  245102  (2002).

\bibitem{nakamura97}
M. Nakamura, K. Nomura, and A. Kitazawa, Phys. Rev. Lett {\bf 79},  3214
  (1997).

\bibitem{white92}
S.~R. White, Phys. Rev. Lett {\bf 69},  2863  (1992).

\bibitem{white93}
S.~R. White, Phys. Rev. B {\bf 48},  10345  (1993).

\bibitem{schollwoeck05}
U. Schollw\"ock, Rev. Mod. Phys. {\bf 77},  259  (2005).

\bibitem{clay99}
R.~T. Clay, A.~W. Sandvik, and D.~K. Campbell, Phys. Rev. B {\bf 59},  4665
  (1999).

\bibitem{ejima05}
S. Ejima, F. Gebhard, and S. Nishimoto, Europhys. Lett {\bf 70},  492  (2005).

\bibitem{kawakami90b}
N. Kawakami and S.-K. Yang, Phys. Lett. {\bf 65},  2309  (1990).

\bibitem{chen96}
L. Chen and S. Moukouri, Phys. Rev. B {\bf 53},  1866  (1996).

\bibitem{assaad91}
F.~F. Assaad and D. W\"{u}rtz, Phys. Rev. B {\bf 44},  2681  (1991).

\bibitem{ogata90}
M. Ogata and H. Shiba, Phys. Rev. B {\bf 41},  2326  (1990).

\bibitem{lavalle03}
C. Lavalle, M. Arikawa, S. Capponi, F.~F. Assaad, and A. Muramatsu, Phys. Rev.
  Lett {\bf 90},  216401  (2003).

\bibitem{pruschke92}
T. Pruschke and H. Shiba, Phys. Rev. B {\bf 46},  356  (1992).

\end{thebibliography}

\end{document}